\documentclass[12pt]{article}

\pdfoutput=1
\usepackage{color}
\usepackage{epsfig, palatino}
\usepackage{pstricks,pst-node,pst-tree}
\usepackage{epic}
\usepackage{mathrsfs}
\usepackage{ae} 
\usepackage[T1]{fontenc}
\usepackage[ansinew]{inputenc}
\usepackage{amsmath}
\usepackage{bbm}
\usepackage{amssymb}
\usepackage{graphicx}
\usepackage{ulem}
\definecolor{darkblue}{cmyk}{0.9,0.9,0,0}
\definecolor{darkgreen}{cmyk}{0.9,0,0.9,0}

\definecolor{blueblue}{cmyk}{0.73,0.28,0,0.5}
\definecolor{lightblue}{RGB}{55,171,200}
\definecolor{grey}{gray}{0.55}

\usepackage[colorlinks=true,linkcolor=darkblue,citecolor=darkblue,urlcolor=darkblue]{hyperref}
\usepackage{cite}
\usepackage{hyperref}
\usepackage{wasysym}
\usepackage{varioref}
\usepackage{makeidx}
\usepackage[english]{babel}
\usepackage{simplewick}
\usepackage{array}
\usepackage{multirow}
\usepackage{tabularx}
\usepackage[font={small}]{caption}

\def\({\left(}
\def\){\right)}
\def\[{\left[}
\def\]{\right]}
\def\<{\langle}
\def\>{\rangle}

\newcommand{\cO}{\mathcal O}

\newcommand{\la}[1]{\label{#1}} 

\newcommand{\beq}{\begin{equation}}
\newcommand{\eeq}{\end{equation}}
\newcommand{\beqq}{\begin{equation*}}
\newcommand{\eeqq}{\end{equation*}}
\newcommand\beqa{\begin{eqnarray}}
\newcommand\eeqa{\end{eqnarray}}

\newcommand{\nn}{\nonumber}

\newcommand{\Green}[1]{{\color{darkgreen}#1\color{black}}}

\newcommand*{\TextVCenter}[1]{%
  \text{$\vcenter{\hbox{#1}}$}%
}

\newcommand{\dpole}{\includegraphics[width=0.1in]{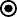}}
\newcommand{\dzero}{\includegraphics[width=0.1in]{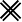}}

\makeindex

\begin{document}

\thispagestyle{empty}

\renewcommand{\thefootnote}{\fnsymbol{footnote}}
\setcounter{page}{1}
\setcounter{footnote}{0}
\setcounter{figure}{0}
\begin{center}
$$$$
{\Large\textbf{\mathversion{bold}
Adding flavour to the S-matrix bootstrap}\par}

\vspace{1.0cm}

\textrm{Luc\'ia C\'ordova$^\text{\tiny 1,2,3}$ and Pedro Vieira$^\text{\tiny 1,4}$}
\\ \vspace{1.2cm}
\footnotesize{\textit{
$^\text{\tiny 1}$Perimeter Institute for Theoretical Physics, Waterloo, Ontario N2L 2Y5, Canada\\
$^\text{\tiny 2}$Department of Physics and Astronomy \& Guelph-Waterloo Physics Institute, University of Waterloo, Waterloo, Ontario N2L 3G1, Canada\\
$^\text{\tiny 3}$ Institut de Physique Th\'eorique, CEA Saclay, 91191 Gif-sur-Yvette, France\\
$^\text{\tiny 4}$
Instituto de F\'isica Te\'orica, UNESP, ICTP South American Institute for Fundamental Research,
01140-070, S\~ao Paulo, Brazil
}  
\vspace{4mm}
}

\par\vspace{1.5cm}

\textbf{Abstract}\vspace{2mm}

\end{center}


We explore the S-matrices of gapped, unitary, Lorentz invariant quantum field theories with a global O($N$) symmetry in 1+1 dimensions. We extremize various cubic and quartic couplings in the two-to-two scattering amplitudes of vector particles. Saturating these bounds, we encounter known integrable models with O($N$) symmetry such as the O($N$) Gross-Neveu and non-linear sigma models and the scattering of kinks in the sine-Gordon model. We also considered more general mass spectra for which we move away from the integrable realm. In this regime we find (numerically, through a large N analysis and sometimes even analytically) that the S-matrices saturating the various coupling bounds have an extremely rich structure exhibiting infinite resonances and virtual states in the various kinematical sheets. They are rather exotic in that they admit no particle production yet they do not obey Yang-Baxter equations. We discuss their physical (ir)relevance and speculate, based on some preliminary numerics, that they might be close to more realistic realistic theories with particle production. 
 
\noindent

\setcounter{page}{1}
\renewcommand{\thefootnote}{\arabic{footnote}}
\setcounter{footnote}{0}

 \def\nref#1{{(\ref{#1})}}

\newpage

\tableofcontents

\parskip 5pt plus 1pt   \jot = 1.5ex

\section{Introduction} 
In this paper we continue the exploration of the space of gapped quantum field theories following  \cite{Paper1,Paper2,Paper3} by focusing on two dimensional theories with a global symmetry under which particles transform. We will consider the two-to-two scattering processes of particles transforming in the vector representation of O($N$). As we will see, such processes turn out to be way richer than their analogous counterparts without global symmetry. 

As described in much more detail in the next section, kinematically, such two-to-two scattering amplitudes live in the Mandelstam physical sheet or equivalently in the physical strip $0<\text{Im}(\theta)<\pi$ in terms of the hyperbolic rapidity $\theta$. Direct $s$-channel processes take place at the lower boundary of this strip, for real rapidity. There, the various amplitudes are bounded by unitarity as $|S_\text{rep}(\theta)|^2\le 1$ where $\text{rep}$ are the various possible representations formed by the two incoming particles. At the upper boundary of the physical sheet we have the crossed $t$-channel processes. Here is where global symmetry manifests itself rather strikingly as a tension between unitarity and crossing symmetry. Indeed, under crossing transformations, the various individual components are trivially swapped  but when we translate that back to the representation basis -- for which $s$-channel unitarity was so conveniently simple -- we obtain a very non-trivial mixing of the various representations. 
In this upper boundary we then have $|\sum_{\text{rep'}}\mathbbm{d}_{\text{rep},\text{rep'}} S_\text{rep'}(i\pi-\theta)|^2\le 1$ where $\mathbbm{d}$ is a constant $N$-dependent matrix of purely group theoretical origin. Exploring the space of $S$-matrices with global symmetry thus amounts to studying the space of functions leaving in this strip and bounded at its boundary in such coupled way. It is this fascinating problem which we will consider here.\footnote{The higher dimensional counterpart is currently being investigated in \cite{Andrea}}

By looking for theories which maximize particular couplings, we will rediscover in this way the two most famous integrable models with O($N$) symmetry -- the O($N$) non-linear sigma model and the O($N$) Gross-Neveu model -- whose S-matrices were found by Zamolodchikov and Zamolodchikov in their seminal 1979 paper \cite{Zam}. For $N=2$, we rediscovered the S-matrix for the kinks of the Sine-Gordon model. Furthermore, we make contact with a much less known integrable solution \cite{newYB}. We also found some other analytic solutions whose physical (ir)relevance is discussed below. Some other times, these maximization problems require numerics. Both in the analytic and numerical examples, we will often unveil a very rich structure of infinitely many  resonances in the various Riemann sheets for the putative S-matrices with the largest possible physical couplings. 



In section \ref{sec2} we introduce the O($N$) S-matrix setup and we review Zamolodchikovs' S-matrices mentioned above. We also explain the general numerical setup in this section. We then discuss in section~\ref{sec3} how the integrable S-matrices can be rediscovered from these numerics when we impose very specific spectra of bound states (or absence thereof). As we move away from these special points, the S-matrices develop very interesting new features. We present in the same section some numerical results for some non-integrable cases. In section~\ref{sec4} we describe an analytic large $N$ analysis which yields some intuition for the rich mathematical structures found numerically. We then present the analytic solution of various cases along with some general analytic properties for the S-matrices. Finally in section~\ref{discussion} we give some concluding remarks.



{\it Note:} While we were finishing the preparation of this article, we became aware of the work \cite{Miguel} which overlaps with section~\ref{sec3} here.

\section{Setup, key examples and numerics}\la{sec2}

\subsection{S-matrices} \la{SmatrixSec}

%
%
%

We consider relativistic particles of mass $m$ transforming in the O($N$) vector representation. Our particles have an O($N$) index $i=1,\dots,N$ and an hyperbolic rapidity $\theta$ parametrizing its energy and momentum as the usual: $E \pm P = m \exp(\pm \theta)$. Lorentz boosts act as translations in hyperbolic rapidity hence Lorentz invariant quantities depend on difference of rapidities only. An example which will be the focus of this paper is the two-to-two particle S-matrix $S_{ij}^{kl}(\theta)$ defined in the standard way\footnote{In two dimensions the initial and final rapidities are the same because of energy-momentum conservation.}
\beq
|\theta_1,i_1;\theta_2,i_2\rangle_\text{out}=\mathbb{S}_{i_1i_2}^{j_1j_2}(\theta=\theta_1-\theta_2) |\theta_1,j_1;\theta_2,j_2\rangle_\text{in} \,. \la{evolution}
\eeq
In terms of Mandelstam invariants, we have $s/m^2=4-t/m^2=2+2\cosh(\theta)$ and $u=0$ in two dimensions. The very useful map between $s$ and $\theta$ opens up the two particle cut and is depicted in figure \ref{fig_sheets}. 

\begin{figure} [t]
\centering
\includegraphics[scale=1.4]{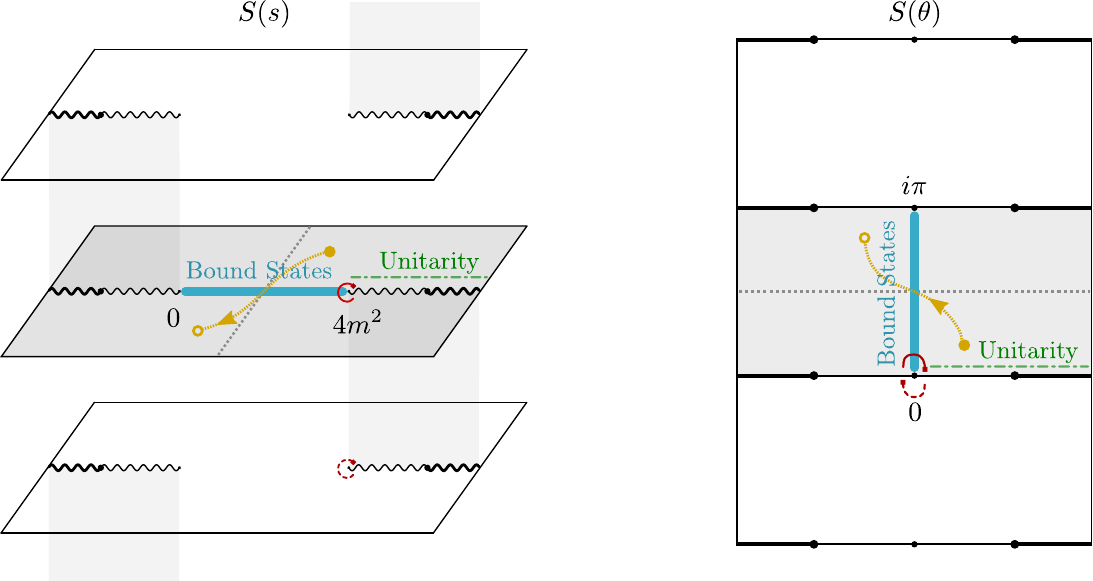}
\caption{Mapping between $s$ and $\theta$ variables. The cuts corresponding to the two particle thresholds with branch points at $s=0$ and $s=4m^2$ get opened in the $\theta$ plane. These cuts are square roots, so that going twice around one of the branch points leaves you back where you began. This is exemplified in the red path for the branch point $s=4m^2$ (or the regular point $\theta=0$). The thick lines in black represent possible inelastic thresholds. Unitarity is imposed for physical values of the center of mass energy $s\geq4m^2$ ($\theta>0$) represented by the green dashed line. The physical sheet(strip) in $s$($\theta$) is highlighted in grey. The bound state poles are located in the window $s\in\[0,4m^2\]$ ($\theta\in\[0,i\pi\]$) (blue line). Finally, the points in yellow are related by crossing which acts as a reflection around $s=2m^2$ ($\theta=i\pi/2$). Notice that in general there are infinitely many sheets.}
\la{fig_sheets}
\end{figure}

The S-matrix can be coded into three functions corresponding to the three possible O($N$) invariant tensors with four indices 
\beq
\mathbb{K}_{ij}^{kl} = \delta_{ij}\delta^{kl} =\vcenter{\hbox{\includegraphics{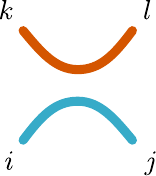}}}\,, \qquad \mathbb{I}_{ij}^{kl} = \delta_{i}^l\delta_j^{k}=\vcenter{\hbox{\includegraphics{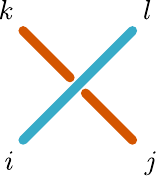}}}\,,\qquad \mathbb{P}_{ij}^{kl} = \delta_{i}^{k}\delta_j^{l}=\vcenter{\hbox{\includegraphics{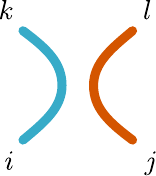}}} \nn
\eeq
or, equivalently, to the three irreducible representations arising in the product of two fundamental incoming representations. Explicitly, we have 
\beqa
\mathbb{S}(\theta)&=&\sigma_1(\theta)\mathbb{K}+\sigma_2(\theta)\mathbb{I}+\sigma_3(\theta)\mathbb{P}\la{SmatrixON}\\
&=&S_{\text{sing}}(\theta)\mathbb{P}_{\text{sing}}+S_{\text{anti}}(\theta)
\mathbb{P}_{\text{anti}}+S_{\text{sym}}(\theta)\mathbb{P}_{\text{sym}} \la{SmatrixON2}\,.
\eeqa
with a trivial translation between these two equivalent descriptions
\beqa
\sigma_{\text{sym}}=\sigma_2+\sigma_3\,,\qquad
\sigma_{\text{anti}}=\sigma_2-\sigma_3\,,\qquad
\sigma_{\text{sing}}=N\sigma_1+\sigma_2 +\sigma_3 \,.\la{rep123}
\eeqa
In the component description crossing symmetry is trivial to impose since under crossing transformations we swap incoming and outgoing particles so that the O($N$) invariant tensors are simply re-shuffled. In the irreducible representation description unitarity is straightforward since by preparing a two-particle state in a definite O($N$) representation we diagonalize~(\ref{evolution}). Crossing symmetry and unitarity in both descriptions are summarized in tables \ref{tab:components} and \ref{tab:channels}.  Alternatively, one could also work in a basis where crossing is diagonalized as in \cite{Martin}.

\begin{table}[t]
 \centering
\begin{tabularx}{\textwidth}{cX}
\hline
 \multicolumn{2}{c} {$S_{ij}^{kl}(\theta)=\sigma_1(\theta)\delta_{ij}\delta^{kl}+\sigma_2(\theta)\delta_i^l\delta_j^k+\sigma_3(\theta)\delta_i^k\delta_j^l$} \\
 \hline
\multirow{2}{4em}{Crossing}
&$\sigma_1(i\pi-\theta)=\sigma_3(\theta)$ \\ 
&  $\sigma_2(i\pi-\theta)=\sigma_2(\theta)$\\
\hline
\multirow{3}{4em}{Unitarity}
& $|\sigma_2(\theta)+\sigma_3(\theta)|^2\leq1$, 
\\ 
& $|\sigma_2(\theta)-\sigma_3(\theta)|^2\leq1$,  \;\; \qquad\qquad\qquad \text{for}\qquad$\theta>0$\\ 
& $|N\sigma_1(\theta)+\sigma_2(\theta)+\sigma_3(\theta)|^2\leq1$\\ 
 \hline
\end{tabularx}
\caption{Crossing and unitarity for the S-matrix in the component decomposition (\ref{SmatrixON}). Crossing simply re-shuffles the functions in this language but unitarity is more complicated.}
\label{tab:components}
\end{table}

We assume here that the particles being scattered are the lightest particles in the theory. There could also be bound states showing up in the S-matrix of two fundamental particles. They would transform in the singlet, anti-symmetric or symmetric traceless representation of O($N$) and show up as a pole in the corresponding channel. For instance, if there is a single bound state transforming in the anti-symmetric representation with mass $m_{\text{BS}}=2 \cos(\lambda/2)$ there is a pole in the corresponding channel as\footnote{$\mathcal{J}=2\sin(\lambda)$ is a trivial Jacobian. Notice that here we define the coupling $g_\text{rep}^{2}$ to be the residue of the S-matrix $S_{\text{rep}}(s)$ and not of $T_{\text{rep}}(s)$ so we would need one further simple Jacobian if we were to compare the results in this paper with the conventions in \cite{Paper2}.}
\beq
S_\text{anti} \simeq \frac{ g_{\text{anti}}^2}{s-m_{\text{BS}}^2} \qquad \Leftrightarrow \qquad S_\text{anti} \sim \frac{\mathcal{J}   g_{\text{anti}}^2}{\theta-i \lambda} \la{eq1}
\eeq
which then, according to crossing, would lead to poles in the t-channel for all components as can be read from table \ref{tab:channels},
\beq
S_\text{sing} \sim (\tfrac{1}{2}-\tfrac{N}{2})\times \frac{\mathcal{J}g_{\text{anti}}^2}{i\pi-\theta-i \lambda}\, , \qquad S_\text{anti} \sim \tfrac{1}{2}\times \frac{\mathcal{J}g_{\text{anti}}^2}{i\pi-\theta-i \lambda} \,,\qquad S_\text{sym} \sim \tfrac{1}{2}\times \frac{\mathcal{J}g_{\text{anti}}^2}{i\pi-\theta-i \lambda} \,. \la{eq2}
\eeq
It is because of the interplay between these various channels, also illustrated in the tension between simplifying unitarity and crossing at the same time as summarized in tables~\ref{tab:components} and~\ref{tab:channels}, that the O($N$) bootstrap problem is so much richer than the single component case.

\begin{table}[t]
 \centering
\begin{tabularx}{\textwidth}{cX}
\hline
 \multicolumn{2}{c} {$\mathbb S(\theta)=S_{\text{sing}}(\theta)\mathbb{P}_{\text{sing}}+S_{\text{anti}}(\theta)
\mathbb{P}_{\text{anti}}+S_{\text{sym}}(\theta)\mathbb{P}_{\text{sym}}$}\\
\hline
\multirow{3}{4em}{Crossing}& $S_{\text{sing}}(i\pi-\theta)=\frac{1}{N}S_{\text{sing}}(\theta)+\(\frac{1}{2}-\frac{N}{2}\)S_{\text{anti}}(\theta)+\(\frac{N}{2}+\frac{1}{2}-\frac{1}{N}\)S_{\text{sym}}(\theta)$\\
&$S_{\text{anti}}(i\pi-\theta)=-\frac{1}{N}S_{\text{sing}}(\theta)+\frac{1}{2}S_{\text{anti}}(\theta)+\(\frac{1}{2}+\frac{1}{N}\)S_{\text{sym}}(\theta)$\\
&$S_{\text{sym}}(i\pi-\theta)=\frac{1}{N}S_{\text{sing}}(\theta)+\frac{1}{2}S_{\text{anti}}(\theta)+\(\frac{1}{2}-\frac{1}{N}\)S_{\text{sym}}(\theta)$\\
 \hline
 Unitarity & $|S_\text{rep}(\theta)|^2\leq1$,  \;\; $\theta>0$\\
 \hline
\end{tabularx}
\caption{Crossing and unitarity for the S-matrix in the representation decomposition (\ref{SmatrixON2}). Unitarity is simple since the various representations diagonalize the scattering evolution while crossing is complicated as it mixes the various representations.
}\label{tab:channels}
\end{table}

\subsection{Integrable O($N$) S-matrices}\la{integrable}

In this section we review the integrable S-matrices with O($N$) symmetry obtained by Zamolodchikov and Zamolodchikov in \cite{Zam}.  In this seminal work, starting from the decomposition \eqref{SmatrixON}, factorized scattering, saturation of unitarity and crossing symmetry are imposed. It was found that under these conditions the S-matrix can be almost completely determined and for N>2 takes the form\footnote{As explained in \cite{Zam}, the Yang-Baxter equations describing factorized scattering have different solutions for $N=2$ and $N>2$. } 
\beq
\mathbf S^\text{int}(\theta)\equiv \begin{pmatrix}S_\text{sing}(\theta)\\S_\text{anti}(\theta)\\S_\text{sym}(\theta)\end{pmatrix}=\prod\limits_{j=1}^{M} \frac{\sinh\theta-i\sin\alpha_j}{\sinh\theta+i\sin\alpha_j}\,\mathbf S^\text{NLSM}(\theta)\,. \la{genSol}
\eeq
where $M$ is a non-negative integer, $\alpha_j$ are free parameters, and  
%
\beq
\mathbf S^\text{NLSM}(\theta)=
\begin{pmatrix}
-1\\
-\frac{\theta-i\pi}{\theta+i\pi} \\
-\frac{\theta-i\pi}{\theta+i\pi}\frac{\theta-i\lambda_\text{GN}}{\theta+i\lambda_\text{GN}}
\end{pmatrix}
F_{\pi+\lambda_\text{GN}}(\theta)F_{2\pi}(\theta)\la{SmatrixNLSM}\, \,, \qquad   F_a(\theta)\equiv \frac{\Gamma \left(\frac{a+i \theta }{2 \pi }\right) \Gamma \left(\frac{a-i \theta +\pi }{2 \pi }\right)}{\Gamma \left(\frac{a-i \theta }{2 \pi }\right) \Gamma \left(\frac{a+i \theta +\pi }{2 \pi }\right)} \,. 
\eeq
where $\lambda_\text{GN}\equiv \frac{2\pi}{N-2}$. Here NLSM stands for Non-linear Sigma Model and GN stands for Gross-Neveu for reasons that will become clear shortly. Let us review and highlight a few of the many very interesting features of (\ref{genSol}) and (\ref{SmatrixNLSM})

\begin{figure}
\centering
\includegraphics[scale=1]{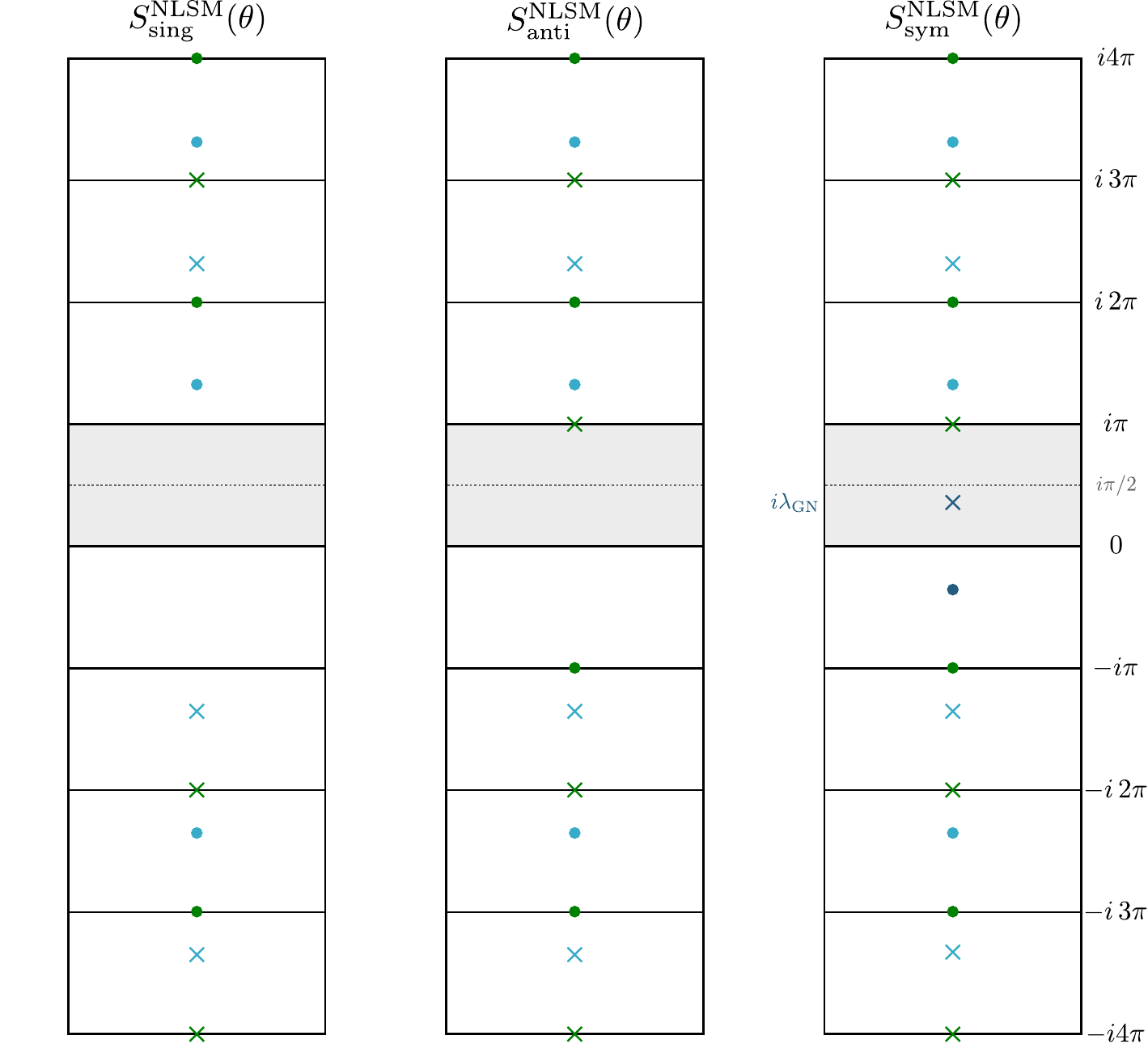}
\caption{Analytic structure of the non-linear sigma model S-matrix in the $\theta$ plane corresponding to the 'minimal' integrable solution. The physical strip is highlighted in grey. The bullets $\color{grey} \bullet$ represent simple poles and the crosses $\color{grey} \boldsymbol\times$ simple zeros. There is a zero ${\color{blueblue}\boldsymbol\times}$ in the physical strip in the symmetric component at the position $\theta=i\lambda_\text{GN}=2\pi/(N-2)$. Integrability implies that unitarity is saturated, i.e. $S_\text{rep}(\theta)S_\text{rep}(-\theta)=1$, so that in a given representation a pole at $\theta$ has an associated zero at $-\theta$. Crossing mixes the different representations relating points at $\theta$ and $i\pi-\theta$ (see discussion around \eqref{eq2}).}
\la{fig_nlsmtheta}
\end{figure}

\begin{enumerate}
\item The first thing to note is that this solution has a very rich pattern of poles and zeros in the infinitely many copies of the Mandelstam $s$ plane (equivalently, in the infinitely many strips in the $\theta$ plane as illustrated in figure \ref{fig_sheets}) as made explicit by the various gamma functions in the functions $F_a(\theta)$. At the same time note that the product $F_{\pi-\lambda_\text{GN}}(\theta)F_{2\pi}(\theta)$ does \textit{not} contain any poles (or zeroes) inside the physical strip. The vector in (\ref{SmatrixNLSM}) multiplying these $F$'s also does not have any poles inside the physical strip (only a zero inside the strip at $\theta= i \lambda_\text{GN}$ and a zero at the boundary of the strip at $\theta=i\pi$ for some components.)
\item Therefore, potential bound state poles must come from the Castillejo-Dalitz-Dyson (CDD) pre-factor which is the product in (\ref{genSol}) for $\alpha_j \in [0,\pi]$. The simplest solution commonly referred to as the 'minimal' solution corresponds to setting $M=0$, i.e. to introducing no CDD factors and hence no bound states.\footnote{There are of course infinitely many other solutions which would also have no bound states. We could for example take any $M>0$ with all $\alpha_j \in [-\pi,0]$ which would correspond to inserting further zeroes in the physical strip but no poles. The 'minimal' solution is minimal because it is not polluted by these extra zeroes and has therefore the simplest analytical structure inside the physical strip.} Beautifully, there is a theory whose S-matrix is precisely given by this choice: it is the O($N$) non-linear sigma model \cite{Zam}. The analytic structure of this solution is presented in figure~\ref{fig_nlsmtheta}.
\item As made explicit in (\ref{genSol}), when there are bound states at generic positions $\theta=i\alpha_j$ (with $\alpha_j \in [0,\pi]$) they are common to all representations (with an important exception to be discussed in the next point). This means that integrable theories produce very degenerate spectra where bound states in different representations come along at once with the same mass. For example, there is no integrable theory where the particles form a single bound state in the singlet channel. (We can view this as a nice feature: the bootstrap of O($N$) symmetric theories with particles in the vector and anti-symmetric representations alone, will necessarily land us outside the integrable world. We will investigate these cases further in section~\ref{numNOint}.) 
\item A simple exception is when we consider a single CDD factor, i.e. $M=1$, with $\alpha_1=\lambda_\text{GN}$. In this case the CDD factor introduces two poles in the physical strip: one at $\theta=i\lambda_\text{GN}$ and another one at $\theta=i\pi-i\lambda_\text{GN}$. However, in the symmetric representation the first one is cancelled by the zero explicitly seen in the vector in \eqref{SmatrixNLSM}. Therefore we are left with an s-channel pole at $\theta=i\lambda_\text{GN}$ for the singlet and anti-symmetric representations and a t-channel pole at $\theta=i\pi-i\lambda_\text{GN}$ for all representations.

The former are identified as the bound states of the theory\footnote{The s- and t-channel poles can be distinguished by the sign of their residue -- see (\ref{poles}) below for the explicit expected signs.}. Since both s-channel poles sit at the same position, the bound states have a common mass $m_\text{sing}=m_\text{anti}=2\,m\cos(\lambda_\text{GN}/2)$. There is also a beautiful physical theory corresponding to this S-matrix: it is the O($N$) Gross-Neveu model  \cite{Zam}. The structure of poles and zeros of this model is depicted in figure~\ref{fig_GNtheta}.

\item There is a subtlety with the last point. For the Gross-Neveu solution the sign of the residues at the s-channel pole $\theta=i \lambda_\text{GN}$ is appropriate for $N=7,8,9,\dots$ but opposite of what it should be for $N=5$ and the pole disappears altogether from the physical sheet as $N=4$ or $N=3$.\footnote{Recall that we are now discussing $N>2$ only; we also skip the $N=6$ case since in this case we have $\lambda_\text{GN}=i\pi-\lambda_\text{GN}$ so that s- and t- channel poles collide and it becomes difficult to distinguish them. } 
\begin{enumerate} 
\item For $N=7,8,9,\dots$ the residues have the proper sign and the bound state interpretation holds perfectly. This case is connected to the $N\to \infty$ limit where things simplify and the theory becomes effectively free.\footnote{Note that the various components of the S-matrix in (\ref{SmatrixON}) go to $1$ or $0$ in this limit as expected for a free theory. However, the singlet channel remains finite since it is a sum of $N$ small quantities of order $\cO(N^{-1})$. In other words, although each element is very small, the phase space is very large leading to a final net result for the singlet channel. This is a general expected feature of large $N$ theories: because of the large phase space, the singlet component should dominate in this limit. We will re-encounter this clearly in a large $N$ analysis in section \ref{largeNsec} below.}
\item For $N\leq4$ the potential bound state poles leave the physical strip. Actually, for $N=3,4$ the Gross-Neveu and NLSM solutions coincide.\footnote{It is simple to see that the CDD factor with $\alpha_1=\lambda_\text{GN}=2\pi/(N-2)$ equals one for those values of N.} This small puzzle is resolved by realizing that the the Gross-Neveu spectrum includes kinks\footnote{These transform in the spinor representation of O($N$) and were studied in \cite{Witten:1977xv,Shankar:1978rb}. The Gross-Neveu spectrum includes as well a tower of anti-symmetric tensors whose S-matrices were obtained in \cite{Karowski:1980kq}.} which are the only stable particles below $N=5$. So for $N=3,4$ the S-matrix above cannot describe vector particles of the Gross-Neveu model since they do not exist at all for these values of $N$.
\item For $N=5$ the poles have opposite signs and must be reinterpreted. Actually, the bound states disappear from the spectrum (see equation (6.11) in \cite{Zam}) so that the poles describe now a pair of kinks instead and correspond to Coleman-Thun poles.
\end{enumerate}
\item Note that at large energy (i.e. $s\to \infty$ or $\theta\to \infty$) we have (for any representation) $S^{\text{int}}_{\text{rep}} \simeq 1 +i a_\text{rep}/\log(s)$ so that at very high energy the S-matrices become effectively free but they approach this free limit very slowly, logarithmically so. This logarithm is quite physical, it is a sign of asymptotic freedom of these O($N$) integrable theories.  \la{lastComment}
\item Finally, there is another solution to the factorization (Yang-Baxter) equations which appeared in the appendix of \cite{newYB} and is much less known. To our knowledge, it is has not been understood if there is a physical theory it corresponds to. We make contact with this solution in section~\ref{sec4}.\footnote{We became aware of the work \cite{newYB} after we published version 1 of the present paper. We thank A.B.~Zamolodchikov and S.~Komatsu for various inspiring discussions of closely related S-matrices which eventually led us to this reference.}
\end{enumerate}

\begin{figure}
\centering
\includegraphics[scale=1]{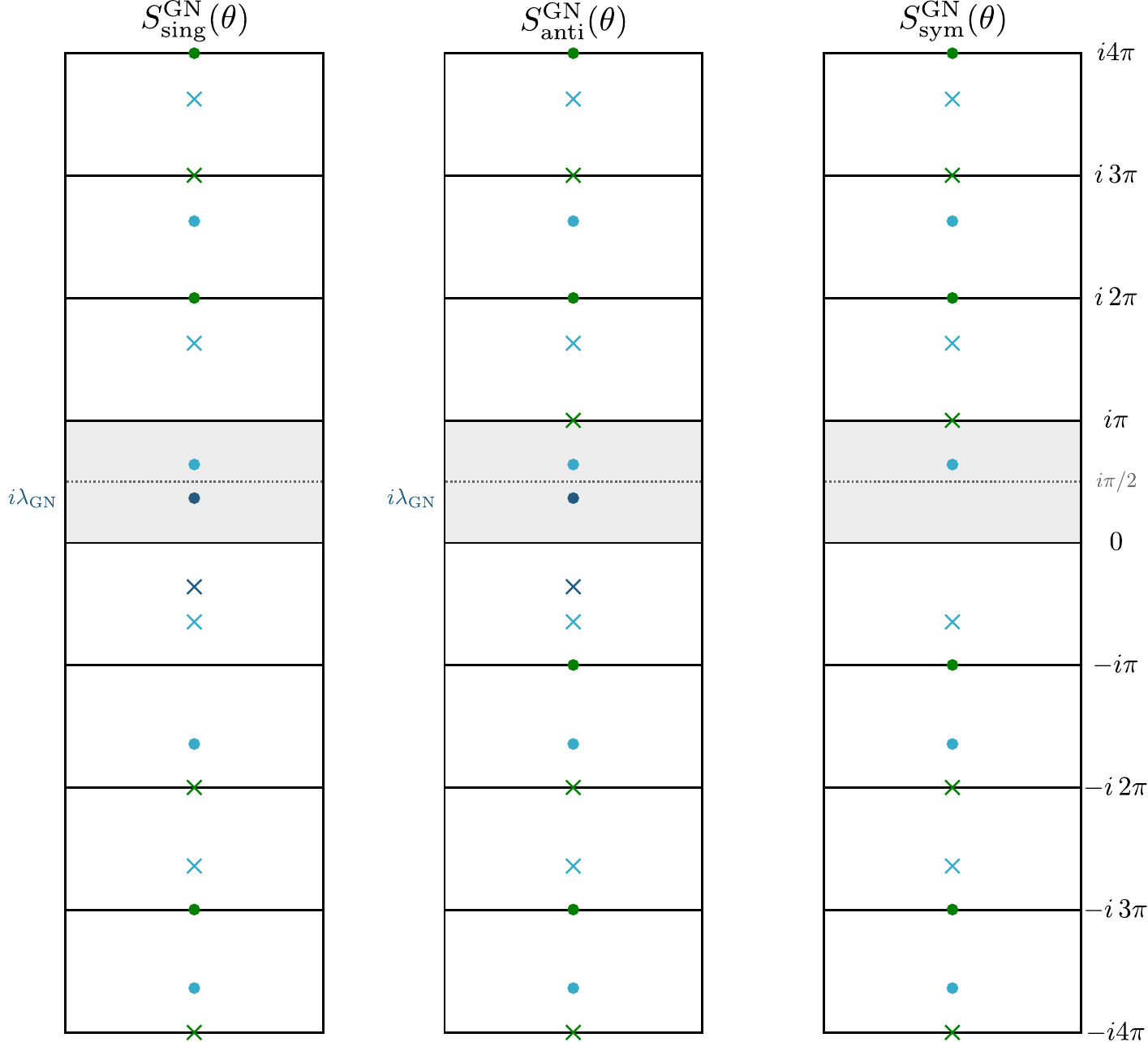}
\caption{Analytic structure of the Gross-Neveu S-matrix in the $\theta$ plane. The conventions are the same as in figure~\ref{fig_nlsmtheta}.  The dark blue poles ${\color{blueblue}\bullet}$ indicate the bound state poles in the singlet and anti-symmetric representations at $\theta=i\lambda_\text{GN}=2\pi/(N-2)$. The lighter blue poles ${\color{lightblue}\bullet}$ depict the t-channel poles at $\theta=i \pi-i\lambda_\text{GN}$ . As explained in the main text, the Gross-Neveu and non-linear sigma model S-matrices are related through a CDD factor which introduces the bound state poles in the singlet and anti-symmetric representations and their corresponding t-channel poles in all representations.}
\la{fig_GNtheta}
\end{figure}

This concludes the review of the $N>2$ solution. For $N=2$ the solution is even richer, with infinitely many gamma functions and is presented in section \ref{sec3int} below. The interpretation of the minimal solution in this case is rather different. It describes the scattering of the kinks and anti-kinks of the sine-Gordon model (which is dual to the massive Thirring model where these kinks correspond to the fundamental fermions). 

\subsection{Numerical setup} \la{NumSec}
Following the analytic structure of the S-matrix, one can write discretized dispersion relations like the ones used in [3] to get numerical bounds on the couplings. In the present case with global symmetry O($N$), there are two main differences: there are three functions $S_\text{rep}(s)$ coupled by crossing and the high energy behaviour requires the use of subtractions.

Crossing symmetry tell us that the vector $\mathbf S=(S_\text{sing},S_\text{anti},S_\text{sym})^\intercal$ satisfies (setting $m=1$ henceforth)
$\mathbf S(4-s)=\mathbbm d\cdot \mathbf S(s)$ where 
\beq
\mathbbm{d}=\left(
\begin{array}{ccc}
\frac{1}{N}&-\frac{N}{2}+\frac{1}{2}&\frac{N+1}{2}-\frac{1}{N}\\
-\frac{1}{N}&\frac{1}{2}&\frac{1}{2}+\frac{1}{N}\\
\frac{1}{N}&\frac{1}{2}&\frac{1}{2}-\frac{1}{N}
\end{array}
\right)\la{matrixd}\,,
\eeq
as can be read off from table~\ref{tab:channels} in section \ref{SmatrixSec}.
Therefore the t-channel poles and the t-channel discontinuities of the vector $\mathbf S$ are related to their s-channel counterparts through this same matrix (while in \cite{Paper2} they were simply identical.) 

As for the slow decay at infinity this is solved by starting our dispersion relation derivation with the identity\footnote{Note that the \textit{subtraction constants} $\mathbf S(2)$ satisfy  the crossing condition $\mathbf S(2)=\mathbbm d\cdot\mathbf S(2)$ which fixes one of the constants in terms of the other two (e.g. we can eliminate $S_{\text{sing}}(2)$ by writing $S_{\text{sing}}(2)=\tfrac{1}{2}[(N+2)S_{\text{sym}}(2)-N S_{\text{anti}}(2)]$). }
\beq
\frac{\mathbf S(s)-\mathbf S(2)}{s-2}= \oint  \frac{dx}{2\pi i} \frac{1}{x-s} \frac{\mathbf S(x)-\mathbf S(2)}{x-2} \la{StartingPoint} \,,
\eeq
where the integral goes over a small circle around a point $s$ inside the physical sheet. Because we divided by $s-2$ the integrand decays very fast at infinity so we can blow up the contour safely\footnote{If the S-matrix approaches a constant at infinity this single subtraction suffices. Note that with this subtraction the integrand vanishes at $1/(x^2 \log(x))$ at infinity for the O($N$) symmetric integrable S-matrices of the previous section. Without the subtraction we would have $\mathbf S(x)/(x-s)\sim 1/(x \log(x))$  which would not decay fast enough to allow us to blow up the contour.} and in this way end up with a contour around the S-matrix poles and the $s-$ and $t-$ channel multi-particle cuts. Using the matrix $\mathbbm d$ in \eqref{matrixd} to relate $t$-channel processes to $s$-channel ones we finally end up with the dispersion relation
\beq
\mathbf S(s)=\mathbf S(2)+\sum_{n=1}^{n_{BS}} \Big(\textbf{pole}_n(s)+ \mathbbm d \cdot \textbf{pole}_n(4-s) \Big)+\int\limits_{4}^{\infty}\frac{(s-2)}{(x-2)}\[\frac{\boldsymbol {\rho}(x)}{x-s} -\mathbbm d\cdot\frac{\boldsymbol \rho(x)}{x-4+s}\]dx \la{subtr} \,,\
\eeq
where the various couplings and bound state masses for the $n_{BS}$ bound states are contained in the vectors
\beq
 \textbf{pole}_n(s)  = \frac{s-2}{m_n^2-2} \frac{g_n^2}{s-m_n^2} ( \delta_{\text{singlet},\text{rep}_n}  ,-  \delta_{\text{anti},\text{rep}_n}  ,  \delta_{\text{sym},\text{rep}_n}  )^\intercal \la{poles}
\eeq
where $\text{rep}_n$ is the representation under which the $n$-th bound state transforms.  For instance, for the non-linear sigma model we have $n_{BS}=0$ so we would need no pole vectors. For the Gross-Neveu model we have $n_{BS}=2$ and these two bound states have the same mass and transform in the singlet and anti-symmetric representation respectively so that in this case we would have two pole vectors $\textbf{pole}^\text{GN}_1(s) = \frac{s-2}{m_\text{GN}^2-2} \frac{g_\text{singlet}^2}{s-m_\text{GN}^2} (1,0,0)^\intercal$ and $\textbf{pole}^\text{GN}_2(s) = \frac{s-2}{m_\text{GN}^2-2} \frac{g_\text{anti}^2}{s-m_\text{GN}^2} (0,-1,0)^\intercal$.

In the dispersion relations \eqref{subtr} we have trivialized crossing taking into account the  O($N$) symmetry of the problem and possible large energy behaviour. The remaining ingredient is to impose unitarity. In order to do this numerically, we follow \cite{Paper2} and choose a grid of $n_\text{grid}$ points $s_{j}>4$ ($j=1,\ldots ,n_\text{grid}$) in which we are to impose unitarity and discretize the densities $\boldsymbol \rho\rightarrow\boldsymbol \rho_j=\boldsymbol \rho(s_j)$ with a linear interpolation between these points so that we can explicitly perform the integrals in \eqref{subtr} and obtain discretized dispersion relations.\footnote{The only difference with respect to \cite{Paper2} is that our integrands have the extra $1/(x-2)$ which slightly changes the expressions as compared to appendix A in \cite{Paper2} in obvious ways.} In this way we impose $n_\text{grid}$ unitarity constraints $|\mathbf S^d(s_j)|\leq1$, where the superscript $d$ denotes the approximated S-matrix using the discretized dispersion relations. 


With this ansatz at hand, we can now choose a bound state spectrum and maximize one of the variables (or combinations of the variables) in the dispersion relation. We could maximize bound state couplings $g_n^2$ as originally proposed in \cite{Paper2} or the value of the S-matrix at a symmetric point $s=t=2$ as in the pion toy models in \cite{Paper3} or we could impose a zero at a given value for some component (i.e. add a resonance) and maximize its slope as in \cite{Doroud}. In the next section we will start with a few such maximization questions (some of which are mixed versions of the ones just mentioned) which lead to the integrable S-matrices discussed previously. Then we will move away from these integrable models and find new S-matrices whose physical (ir)relevance we shall speculate about. 


\section{Numerical results}\la{sec3}
\subsection{Reproducing integrable models}\la{sec3int}
In this section we show how we can reproduce known integrable theories with O($N$) global symmetry. In general, we expect the numerical maximization to reproduce the physical S-matrices for the appropriate mass spectrum. First, we consider a simple problem which reproduces free theory. Then we focus on the case with bound states and $N>2$ which gives rise to the Gross-Neveu S-matrix. After that we discuss the maximization procedure leading to the non-linear sigma model. Finally we present the case with $N=2$ and bound states which reproduces the S-matrix for the kinks of sine-Gordon (or massive Thirring model).

\subsubsection*{Free theory}
Consider a theory with no bound states. Then the various S-matrix components have no poles inside the strip and are therefore bounded by their values at the two boundaries of the strip (i.e. the lower boundary at $\theta\in\mathbb R$ and the upper one at $\theta\in i\pi+\mathbb R$). At the lower boundary of the physical strip we have unitarity which states that in any of the three representations we have $|S_\text{rep}(\theta)|\le 1$ while in the upper boundary these functions can be as large as allowed by the crossing relations in table~\ref{tab:channels} which relate that upper boundary with the value of the same functions in the lower boundary.  For the symmetric component these equations tell us that we can have modulus as large as $1$ in the upper boundary which is the same bound as in the lower boundary.\footnote{For the other components the analysis is more complicated and indeed below we will analyze these cases and obtain a myriad of very rich S-matrices. For instance from the first relation in table~\ref{tab:channels} we learn that the singlet component can have magnitude as large as $N$ in the upper boundary with the bound saturated iff the various components saturate unitarity for real $\theta$ and are appropriately aligned phases. Similarly, from the second relation we lean that the anti-symmetric component can have absolute value as large as $1+2/N$ in the upper boundary.  } Therefore the largest this component can be, anywhere inside the strip is $1$ and this is attained for $S_\text{symmetric}=1$ which in turn leads to $S_{\text{rep}}=1$ for all representations. We see that the theory with no bound states and the largest absolute value for the symmetric component anywhere in the physical strip is the free theory. This is something we can check numerically; it obviously works and is indeed a nice albeit trivial check of our code. 

\subsubsection*{Gross-Neveu}
\begin{figure}[t]
\centering
\includegraphics[width=\textwidth]{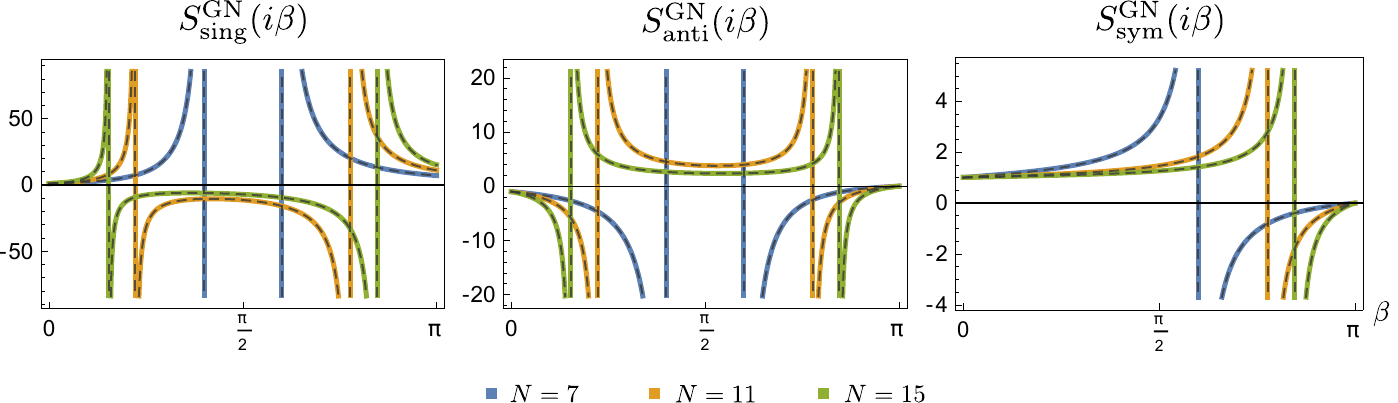}
\caption{Numerical results for the Gross-Neveu spectra $m_\text{GN}=2\cos\(\frac{\pi}{N-2}\)$ maximizing the anti-symmetric coupling $g_\text{anti}$ in the $\theta=i\beta$ line.  Notice that the singlet (left) and anti-symmetric (center) S-matrices have two poles: one for each bound state at $\theta=i\lambda_\text{GN}=i2\pi/(N-2)$ and another one imposed by the crossing equations at $\theta=i\pi-i\lambda_\text{GN}$. The symmetric channel (right) has only the latter pole. The three curves in each representation correspond to different values of N: 7 (blue), 11 (orange) and 15 (green). The bound state pole approaches $\theta=0$ ($s=4m^2$) as we increase the parameter $N$.
The numerical results are in perfect agreement with analytic solutions \eqref{SmatrixGN} plotted in dashed lines. The results were obtained with $n_\text{grid}=70$.}
\la{fig_GNnum}
\end{figure}
As described in section~\ref{integrable}, there are two bound states appearing in the $2\rightarrow2$ S-matrix of the fermions of the Gross-Neveu model. These bound states appear in the {\bf singlet and anti-symmetric representations} and have a common mass $m_\text{GN}=2\cos\(\frac{\pi}{N-2}\)$.
In principle we could maximize any of the two corresponding cubic couplings $g_\text{sing},g_\text{anti}$, but it turns out that it is the maximization of $g_\text{anti}$ which reproduces the Gross-Neveu S-matrix.

Considering the spectrum above in the dispersion relations \eqref{subtr} and implementing the numeric maximization of $g_\text{anti}$ as explained in section~\ref{NumSec} we find perfect agreement with the integrable solution
\beq
\mathbf S^\text{GN}(\theta)=
\begin{pmatrix}
\frac{\theta+i\pi}{\theta-i\pi}\frac{\theta+i\lambda_\text{GN}}{\theta-i\lambda_\text{GN}}\\
\frac{\theta+i\lambda_\text{GN}}{\theta-i\lambda_\text{GN}}\\
1
\end{pmatrix}
F_{\pi-\lambda_\text{GN}}(\theta)F_{\pi}(-\theta)\la{SmatrixGN}\,,
\eeq
where $F_a(\theta)$ is defined in \eqref{SmatrixNLSM}. We performed the numerics for various values of $N$. In figure~\ref{fig_GNnum} we show some numerical results against the solution \eqref{SmatrixGN} for N=7,11,15. The analytic structure of one of the representations is depicted in figure \ref{fig_sheetsGN}.

\begin{figure}[th!]
\centering
\includegraphics[scale=1.35]{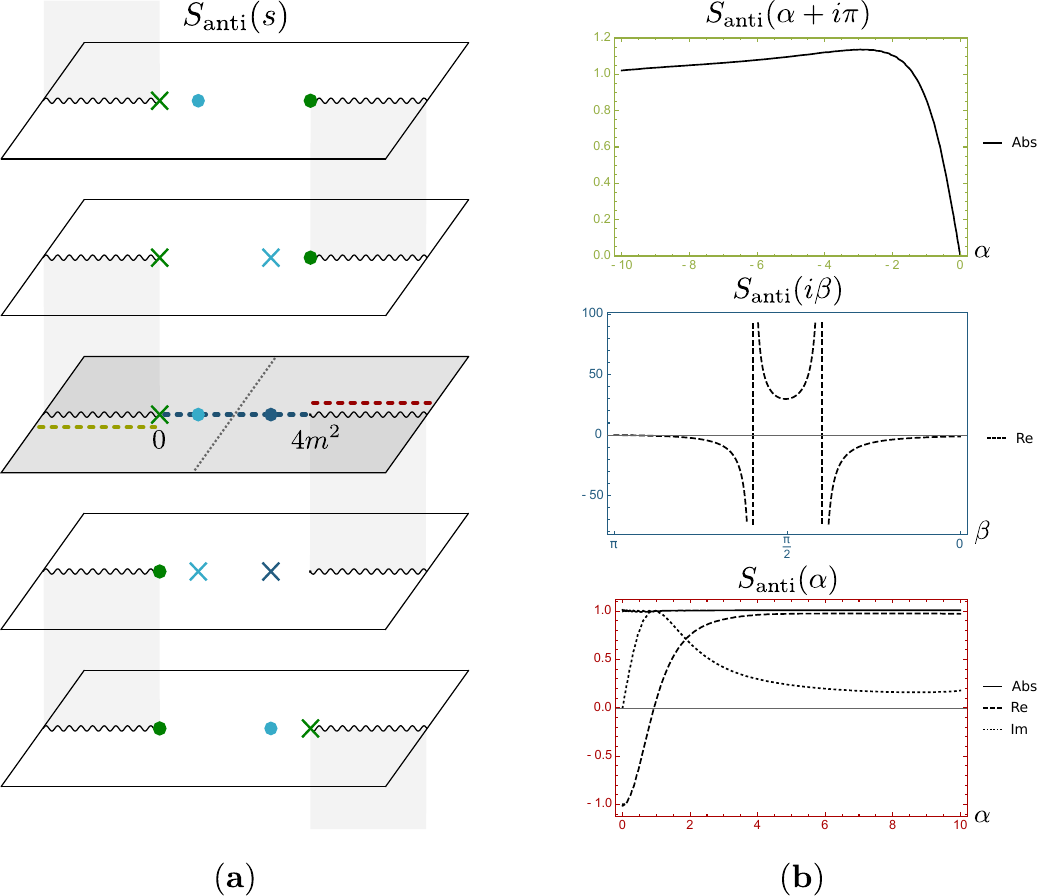}
\caption{
({\bf a}) Analytic structure of the Gross-Neveu anti-symmetric S-matrix in the Mandelstam plane. There is a zero at $s=0$ (in green), a bound state pole at $s=m_\text{GN}^2$ (in dark blue) and a t-channel pole in lighter blue. The green, blue and red dashed lines are respectively the regions below the left cut, between $s\in[0,4m^2]$ and above the right hand cut where we impose unitarity. ({\bf b}) Numerical results for the corresponding regions in the $\theta=\alpha+i\beta$ plane wit $N=7$ ($m_\text{GN}=2\cos(\pi/5)$) and $n_\text{grid}=75$. The top panel in green depicts the zero at $\theta=i\pi$ ($s=0$). The middle panel in blue shows the bound state and t-channel poles. In the red bottom panel we see that unitarity is saturated (solid curve representing the absolute value of the function).
}
\la{fig_sheetsGN}
\end{figure}

\subsubsection*{Non-linear sigma model}
Another famous integrable model we would like to reproduce is the non-linear sigma model. As discussed in section~\ref{integrable}, this model has no bound states and its S-matrix differs from the Gross-Neveu one by an overall CDD factor. Instead of bound states, the parameter $\lambda_\text{GN}$ in \eqref{SmatrixNLSM} labels the position of a zero in the symmetric representation. In the numerics, we impose this zero $S_\text{sym}(\theta=i\lambda_\text{GN})=0$ as well as the absence of bound states by setting all cubic couplings to zero. Inspired by the Gross-Neveu case, we maximize the effective quartic anti-symmetric coupling given by $S_\text{anti}(i\pi/2)$. With these conditions we are able to reproduce the solution \eqref{SmatrixNLSM} as exemplified in figure~\ref{fig_NLSMnum} for $N=6$. 

\begin{figure}
\centering
\includegraphics[scale=1]{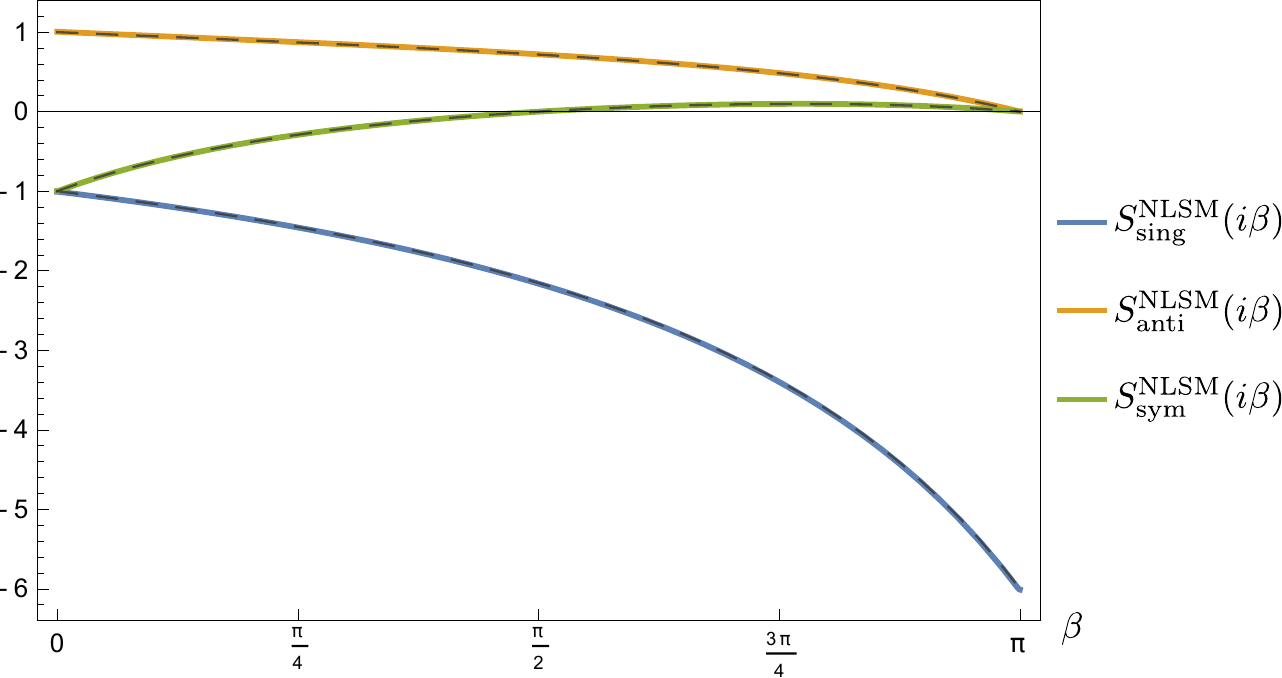}
\caption{Numerical results imposing the NLSM zero at $S_\text{sym}(i\lambda_\text{GN})$ and maximizing $S_\text{anti}(i\pi/2)$ for $N=6$ and $n_\text{grid}=40$. The singlet, anti-symmetric and symmetric S-matrices are plotted respectively in blue, orange and green. These curves perfectly agree with the solution \eqref{SmatrixNLSM} in dashed lines. For this value of $N$ the zero sits exactly at the middle of the strip at $\theta=i\pi/2$.
Similar plots are obtained for any $N\geq5$ where only the position of the latter zero changes.}
\la{fig_NLSMnum}
\end{figure}
As we decrease $N$, the zero in the symmetric representation moves towards the upper boundary of the physical strip, reaching it at $N=4$ and leaving the physical strip for $N<4$. It would be interesting to understand what is the condition one should impose in order to reproduce the non-linear sigma model for $N<5$.

In the beautiful work \cite{Martin} it was pointed out that the space of unitarity, crossing symmetric relativistic S-matrices with O($N$) symmetry has a very rich geometric structure with special cusps. Provided we point more or less towards one such cusp, any maximization functional will push us there. They could identify the non-linear sigma model as a cusp in this theory space, without imposing the zero at $S_\text{sym}(\theta=i\lambda_\text{GN})$. It would be fascinating to explore more throughly the geometry of this S-matrix space, in particular for the various other maximization problems considered in the following sections.


\subsubsection*{Sine-Gordon kinks}

\begin{figure}
\centering
\includegraphics[width=.7\textwidth]{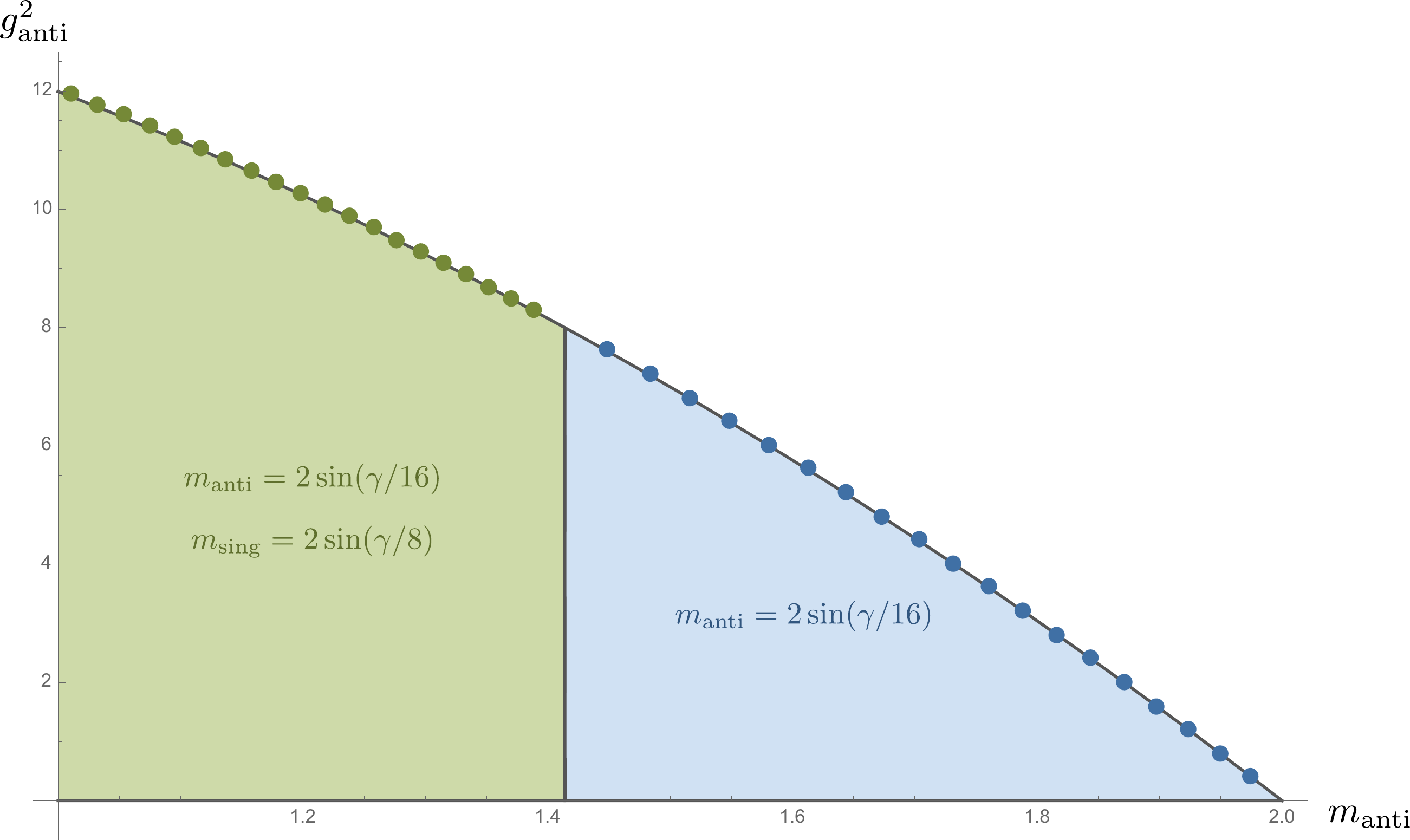}
\caption{Comparison between the $N=2$ sine-Gordon kinks solution and numerical results. In the blue region corresponding to $4\pi<\gamma<8\pi$ we impose one bound state in the anti-symmetric representation with mass $m_\text{anti}=2\sin(\gamma/16)$ and a zero $S_\text{sing}[i(-\pi+\gamma/8)]=0$. In the green region where $8\pi/3<\gamma<4\pi$  we keep the anti-symmetric bound state and impose a new bound state in the singlet representation with mass $m_\text{sing}=2\sin(\gamma/8)$ as well as a zero $S_\text{sing}[i(-\pi+\gamma/2)]=0$. (We avoid the point $m_\text{sing}=\sqrt 2$ because of our subtraction scheme.) The numerical points obtained with $n_\text{grid}=40$ match perfectly with the analytic solution (truncating the infinite product in \eqref{U_N2} to 9) depicted in the grey curve. }
\la{fig_N2kinks}
\end{figure}

Another important result derived in \cite{Zam} is given by the case $N=2$ corresponding to the integrable S-matrix for the sine-Gordon kinks and antikinks.\footnote{Equivalently, one can think about the fundamental fermions of the massive Thirring model.} The kink $A$ and antikink $\bar A$ can be packed into a doublet $A=A_1+iA_2$ ($\bar A=A_1-iA_2$). The solution has a free parameter $\gamma$ \footnote{As explained in \cite{Zam}, this parameter is related to the coupling in the sine-Gordon Lagrangian $\mathcal L_\text{sG}=\frac{1}{2}(\partial\phi)^2+m_0^2/\beta^2\cos(\beta\phi)$ through $\gamma=\beta^2/(1-\frac{\beta^2}{8\pi})$.} and reads
\beq
{\bf S}(\theta)=-\frac{1}{\pi}\,U(\theta)
\begin{pmatrix}
\sin\(\frac{8\pi i\theta}{\gamma}\)-\sin\(\frac{8\pi^2}{\gamma}\)\\ 
\sin\(\frac{8\pi i\theta}{\gamma}\)+\sin\(\frac{8\pi^2}{\gamma}\)\\ 
-\sin\(\frac{8\pi(\pi+i\theta)}{\gamma}\)\\
\end{pmatrix}\,,\la{SmatrixKinks}
\eeq
where the prefactor is given by
\beqa
U(\theta)&=&\Gamma\(\frac{8\pi}{\gamma}\)\Gamma\(1+i\frac{8\theta}{\gamma}\)\Gamma\(1-\frac{8\pi}{\gamma}-i\frac{8\theta}{\gamma}\)\prod\limits_{n=1}^\infty\frac{R_n(\theta)R_n(i\pi-\theta)}{R_n(0)R_n(i\pi)}\la{U_N2}\,,\\
R_n(\theta)&=&\frac{\Gamma\[2n\frac{8\pi}{\gamma}+i\frac{8\theta}{\gamma}\]\Gamma\[1+2n\frac{8\pi}{\gamma}+i\frac{8\theta}{\gamma}\]}{\Gamma\[(2n+1)\frac{8\pi}{\gamma}+i\frac{8\theta}{\gamma}\]\Gamma\[1+(2n-1)\frac{8\pi}{\gamma}+i\frac{8\theta}{\gamma}\]}\nn\,.
\eeqa

The S-matrix \eqref{SmatrixKinks} has a very rich structure. As we vary the parameter $\gamma$, the bound state spectrum changes. The bound state masses are given by $m_n=2m\sin(n\gamma/16)$ with $n<8\pi/\gamma$ and they alternate between the anti-symmetric ($n$ odd) and singlet representations ($n$ even). The bound state with $n=1$ is the sine-Gordon breather whose S-matrix was rediscovered in \cite{Paper2}. For $\gamma\geq8\pi$ all bound states disappear from the spectrum (i.e. the corresponding poles leave the physical strip).

We shall focus on two regimes which comprehend $1<m_1<2$:

${\color{lightblue}\bullet}$ For $4\pi\leq\gamma<8\pi$ there is a single bound state in the {\bf anti-symmetric representation} whose mass takes values $\sqrt2<m_\text{anti}<2$ and is given by $m_\text{anti}=2\sin(\gamma/16)$. 
However, imposing this bound state and maximizing $g_\text{anti}$ is not enough to reproduce the integrable model. Taking inspiration from the NLSM discussed in the previous section, we shall impose one of the zeros in the solution \eqref{SmatrixKinks}.  One zero which is always inside the physical strip for this range of $\gamma$ is  $S_\text{sing}[i(-\pi+\gamma/8)]=0$ which can easily be seen in the first component of~\eqref{SmatrixKinks}.

$\Green{\bullet}$ For $8\pi/3<\gamma<4\pi$ there is another bound state in the  {\bf singlet representation} with mass $m_\text{sing}=2\sin(\gamma/8)$, as well as the previous bound state which for this range takes values $1<m_\text{anti}<\sqrt2$. For the numerics, we keep maximizing the anti-symmetric coupling and impose one zero inside the physical strip, this time at $S_\text{sing}[i(-\pi+\gamma/2)]$.

In figure~\ref{fig_N2kinks} we show the maximum coupling $g_\text{anti}$ obtained numerically for both regimes which nicely matches the analytic solution \eqref{SmatrixKinks}. The S-matrices agree perfectly as well. 

In \cite{Paper2} it was shown that the sine-Gordon  S-matrix for the lightest breathers is quite special. It has a single bound state flowing (the second lightest breather) and of all theories with a single bound state it is the one with the largest possible coupling. Here we see that the kink S-matrix of sine-Gordon also comes about from a maximization procedure as the theory with $O(2)$ symmetry with the largest possible coupling and appropriate resonances (zeroes). These two are related. As mentioned earlier, breathers are themselves bound states of kinks so we can obtain the breather S-matrix by fusing the kink S-matrix. In that sense, the kink S-matrix is the most fundamental one and it is nice to see it come up naturally here. There are also intermediate objets such as the scattering of kinks and breathers. It would be very interesting to look for those as well.

\subsection{Away from integrable points}\la{numNOint}

\begin{figure}
\centering
\includegraphics[scale=0.9]{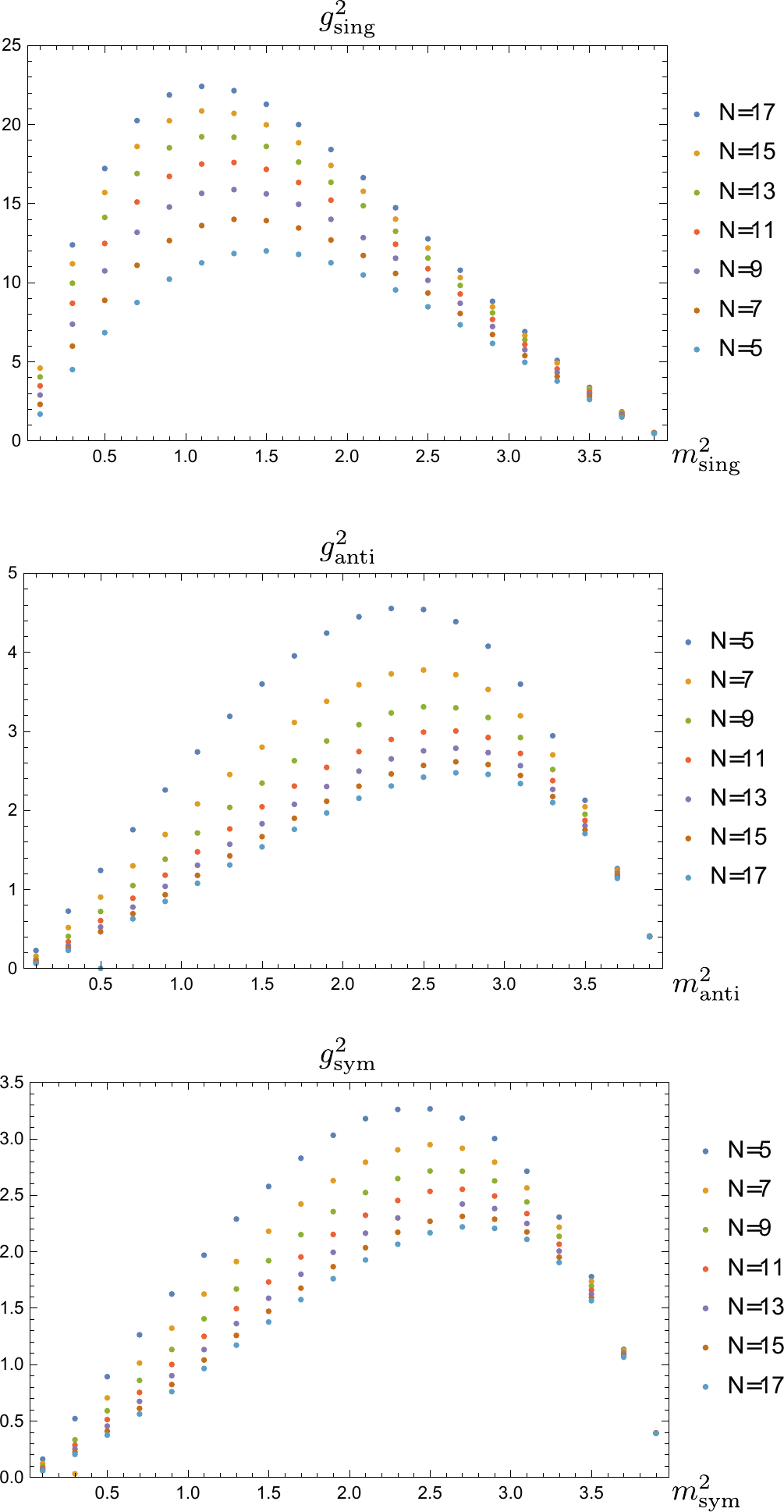}
\caption{Maximum couplings obtained numerically for the case of a single bound state in the singlet (top), anti-symmetric (middle) or symmetric (bottom) representation. The three panels correspond to the three possible representations.  (The numerics were done with $n_\text{grid}=70$ and various $N$'s.)}
\la{fig_onebs}
\end{figure}

In this section we present some of the numerical results obtained for non integrable spectra. The simplest possibility is to consider theories with a single bound state transforming in one of the three representations and maximize its associated cubic coupling $g_\text{rep}$. Figure~\ref{fig_onebs} depicts the maximum couplings as we vary the bound state mass $m_\text{rep}$ and $N$. In these numerics, we can also read of the corresponding S-matrices. We observe -- as in \cite{Paper2} -- that they all saturate unitarity (and thus admit no multi-particle production). That is, for~$s>4m^2$ they all satisfy $|S_\text{rep}|=1$ within our numerical precision. Also as in \cite{Paper2} we observe that these couplings vanish as the bound state becomes weakly coupled as expected. It should be possible (and interesting) to analyse this perturbative corner analytically.\footnote{The interplay with no-particle production could be very interesting to study here since there are no theories with a simple perturbative scalar Lagrangian, O($N$) symmetry and no particle production \cite{Gabai}. The known integrable models with O($N$) symmetry are strongly coupled isolated points.} In the next section we will show how these maximal coupling curves (and others) can be reproduced analytically at large $N$. 

What is \textit{not} shown in figure \ref{fig_onebs} is how interesting the corresponding S-matrices are. By plotting them in the complex $\theta$ plane we found remarkably rich structures of infinite poles and zeros in the various $\theta$-strips.\footnote{Note that we explore the S-matrices across all the Mandelstam cuts while our numerical ansatz is originally designed to parametrize the physical sheet alone. We can do this because unitarity is always saturated from $4m^2$ until the first inelastic threshold which leads to an exact functional relation $S_\text{rep}(\theta)S_\text{rep}(-\theta)=1$ which we can combine with crossing to visit any strip in $\theta$, i.e. any sheet in $s$. See e.g. \cite{Z2} for a similar analysis for a theory with a single particle.} An example is depicted in figure \ref{fig_sheets1bsSING}.

Interesting as they might be, we will not be able to compare the theories saturating the bounds in figure \ref{fig_onebs} with any known theory. The reason is that since they have no particle production, they should probably be integrable. But since they have a single bound state they cannot satisfy Yang-Baxter. So, if they exist they must be something more exotic. We will speculate further on the physics of such theories in the discussions.

\begin{figure}[th!]
\centering
\includegraphics[scale=1.35]{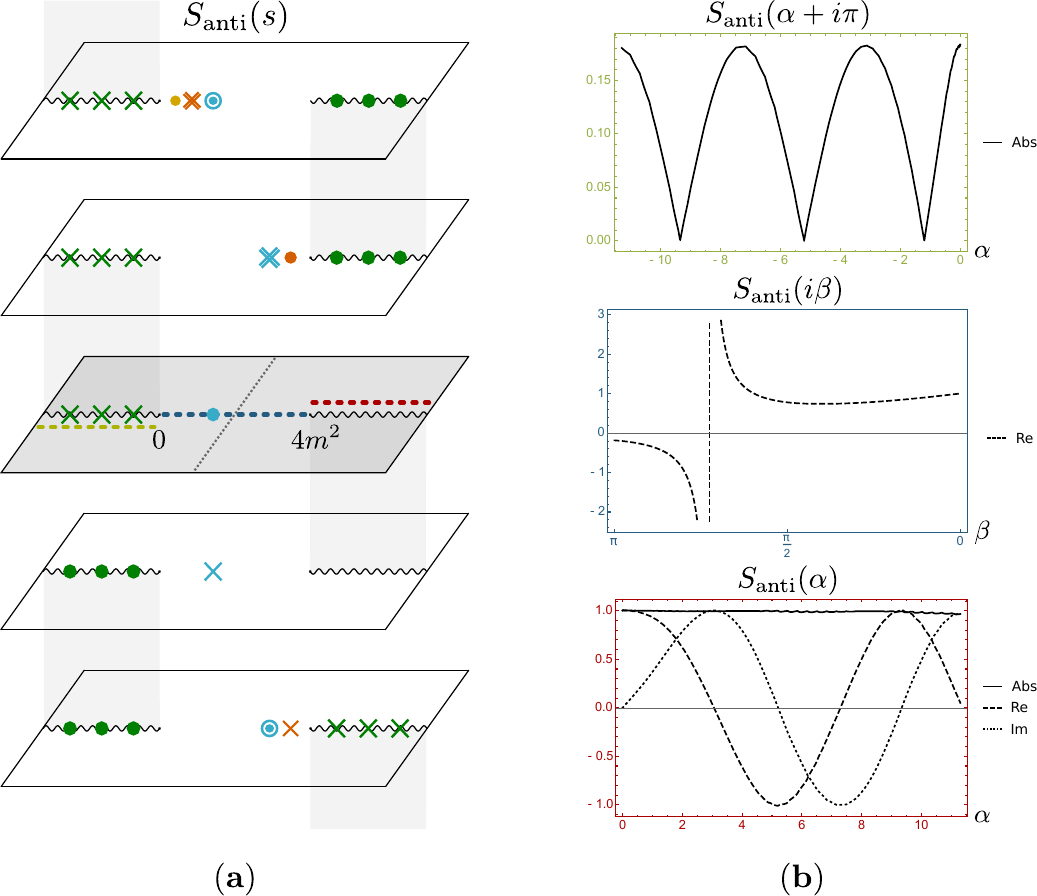}
\caption{({\bf a}) Analytic structure for $S_\text{anti}(s)$ for a theory with a single singlet bound state with maximal cubic coupling. Note the presence of zeroes in the left cut of the physical sheet and the t-channel pole depicted in blue. The red line above the right cut is where we impose unitarity. ({\bf b}) Numerical results for this case in various regions of the $\theta=\alpha+i\beta$ plane with $m_\text{singlet}=\sqrt{3}$, $N=7$ and $n_\text{grid}=70$. In the top panel in green we see that there are infinitely many zeros (resonances) in the $\theta=\alpha+i\pi$ line corresponding to the left cut in the Mandelstam plane. The middle panel in blue shows the t-channel pole. In the red bottom panel, the solid line (absolute value of the function) exhibits the saturation of unitarity, whereas the dashed and pointed lines (real and imaginary parts) show the oscillatory behaviour leading to the zeros of the top green panel. This very rich analytic structure is to be compared with the much simpler figure \ref{fig_sheetsGN} encountered before!
}
\la{fig_sheets1bsSING}
\end{figure}

To make contact with known integrable examples we can consider one bound state in the anti-symmetric representation and another one in the anti-symmetric representation with equal mass and maximize the cubic coupling to the anti-symmetric one. When the mass passes by the value $m_\text{GN}=2\cos(\tfrac{\pi}{N-2})$ we do obtain the Gross-Neveu model as seen in the previous section. Away from this point, the bound is deformed smoothly but the S-matrices are quite different, with a very exotic analytic structure akin to that illustrated in figure \ref{fig_sheets1bsSING}. The maximum couplings we obtained for various $N$ and bound state masses are shown in figure \ref{fig_GNline}. The divergence at $m=\sqrt{2}$ comes from the collision of $s$ and $t$-channel poles as in \cite{Paper2}. In this case they screen each other by simply setting~$g_\text{sing}^2=\frac{N}{2} g_\text{anti}^2$. 

Below, we will revisit some of the cases found numerically here. To do so, first we will get some further insight from a large $N$ analysis in section \ref{largeNsec} before attacking the full finite $N$ case in section \ref{finiteNsec}.


%
%
%

\begin{figure}
\centering
\includegraphics[scale=1]{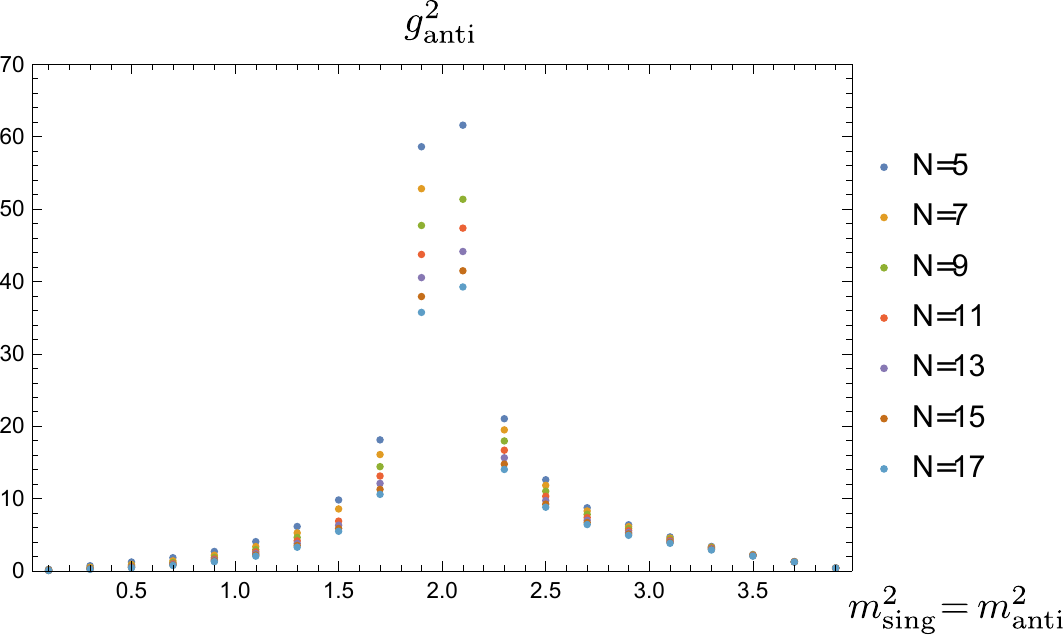}
\caption{Maximum couplings obtained numerically for the case of one bound state in the anti-symmetric representation and another one in the anti-symmetric representation with equal mass. We maximize the anti-symmetric coupling (as in the Gross-Neveu model) and plot for various $N$. The maximum coupling decreases for larger $N$. The numerics were done with $n_\text{grid}=70$.}
\la{fig_GNline}
\end{figure}

\section{Analytic results}\la{sec4}

\subsection{Large $N$} \la{largeNsec}
In this section we will consider large $N$. We will be able in this way to proceed analytically to a much greater extent and get some insights about the general analytic structures to be expected in the finite $N$ optimization numerics. 

The key place where $N$ enters and where large $N$ presents some simplifications is in crossing represented in table \ref{tab:channels} and quoted here again for convenience:
\beqa
S_{\text{sing}}(i\pi-\theta)&=&
\frac{1}{N}S_{\text{sing}}(\theta)+\(\frac{1}{2}{\color{blue}-\frac{N}{2}}\)S_{\text{anti}}(\theta)+\({\color{blue}\frac{N}{2}}+\frac{1}{2}-\frac{1}{N}\)S_{\text{sym}}(\theta) \la{C1}\\
S_{\text{anti}}(i\pi-\theta)&=&-\frac{1}{N}S_{\text{sing}}(\theta)+\frac{1}{2}S_{\text{anti}}(\theta)+\(\frac{1}{2}+\frac{1}{N}\)S_{\text{sym}}(\theta) \la{C2}\\
S_{\text{sym}}(i\pi-\theta)&=&\frac{1}{N}S_{\text{sing}}(\theta)+\frac{1}{2}S_{\text{anti}}(\theta)+
\(\frac{1}{2}-\frac{1}{N}\)S_{\text{sym}}(\theta) \la{C3}
\eeqa
At real $\theta$ all S-matrices saturate unitarity; they are phases so they are all order $\cO(N^0)$ functions, even at large $N$. At the upper boundary of the physical strip, at $\theta \in i\pi+\mathbb{R}$, we can use these crossing equations to see what values the various S-matrices can take. If we can bound the functions in the two boundaries of the physical strip then we can use the maximum modulus principle to bound the functions everywhere inside the strip. That is the main idea which we will explore. 

The singlet component can be very large in the upper boundary of the strip. It can be of order $\cO(N)$ since it is a linear combination of phases and two of them are weighted by huge coefficients, highlighted in blue in \eqref{C1}. The anti-symmetric and symmetric components will be at most of order $\cO(N^0)$ at the upper boundary since they are linear combinations of phases but the prefactors in~(\ref{C1}-\ref{C2}) are at most of order $\cO(N^0)$. We can imagine the values inside the strip to smoothly interpolate between their two boundaries. Then we would conclude that anti-symmetric and symmetric representations would remain small while singlet would be the much larger dominant contribution. This is of course what we expect in a large $N$ theory since there is a lot of phase space to scatter into when we form a singlet. Technically, in large $N$ vector theories, dimmer-like colour loops dominate. Indeed, in all the examples we will now explore, it is by analyzing the singlet component that we can learn a great deal very easily. 

\subsubsection{Maximum singlet effective quartic coupling}\la{max_sing}

Suppose we are after the S-matrix with the largest singlet component in the middle of the physical strip, that is at $\theta= i\pi/2$, in a theory with no bound states (i.e. no poles inside the physical strip). The value at the middle of the strip can only be as large as its value at the boundary of the strip by the maximum modulus principle. At the upper boundary, the singlet S-matrix can have at most absolute value $N/2+N/2=N$ if the phases $S_\text{sym}$ and $S_\text{anti}$ are opposite to each other (at least for real $\theta$) so that the terms in blue in (\ref{C1}) add up. So all in all, we have a function without poles which satisfies $|S_\text{singlet}(\theta)|\le 1$ at real $\theta$ and $|S_\text{singlet}(i\pi+\theta)|\le N$ at the other boundary of the physical strip. One function which clearly saturates these bounds is
\beq
S_\text{singlet}^\text{optimal}(\theta)=e^{ \frac{\theta}{i \pi} \log(N)} \la{singletLargeN}
\eeq
which is indeed a phase for $\theta \in \mathbb{R}$, equal to $N$ times a phase for $\theta\in i\pi + \mathbb{R}$ and real analytic for purely imaginary $\theta$. This is the optimal solution to our problem since the ratio $S_\text{singlet}\theta)/S_\text{singlet}^\text{optimal}(\theta) $ has absolute value smaller or equal to $1$ at both boundaries and thus is at most $1$ in the middle of the strip. Hence 
\beq
\text{Max} [S_\text{singlet}(i\pi/2)]= \sqrt{N} \,. \la{largeNMiddle}
\eeq
Now we can test this against the numerics described before. Performing them for various values of $N$ leads to results in figure \ref{figSsingletMiddleStrip} which beautifully matches with our prediction (\ref{largeNMiddle}) as we extrapolate to infinite $N$. 
\begin{figure}
\centering
\includegraphics[scale=0.8]{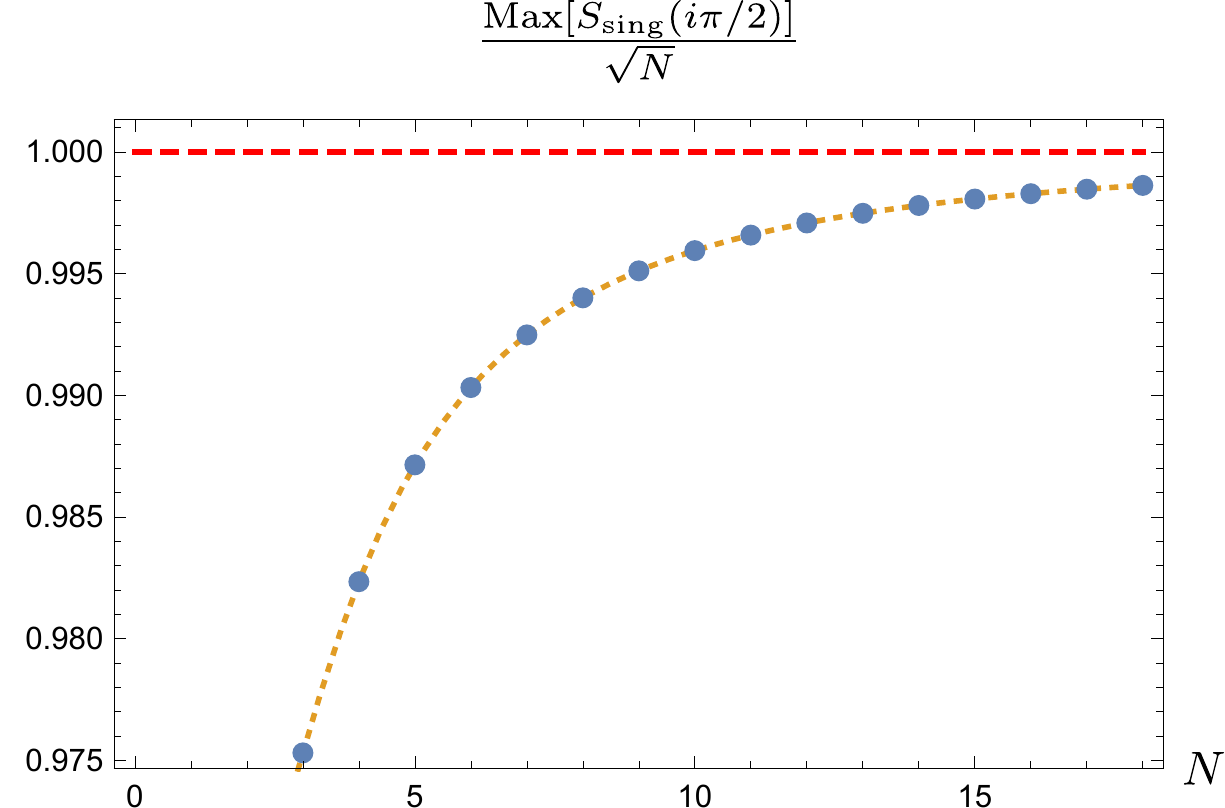}
\caption{Numerical results of the maximization procedure described in section \ref{NumSec} for the maximal value of the singlet S-matrix at the middle of the physical strip for various values of $N$. We expect the result to approach $\sqrt{N}$ as $N\to \infty$ as illustrated by the red dashed line. The yellow dotted line which perfectly passes by the numerical blue dots is the result of a more refined finite $N$ analytic analysis which we will consider in section~\ref{finitemaxSING}.}
\la{figSsingletMiddleStrip}
\end{figure}

We can also learn about all the S-matrices themselves and their analytic properties. As we saw before, for the singlet component to be as large as possible the symmetric and anti-symmetric components ought to be opposite of each other  for real $\theta$ so that the two blue terms in (\ref{C1}) would add up constructively. In this case, where we impose no poles that force these components to be different, it is natural to assume that $S_{\text{sym}}=-S_{\text{anti}}$ everywhere. (This is confirmed by the numerics at any finite $N$.) Then, (\ref{C1}) immediately leads to
\beqa
S_{\text{sym}}(\theta)&\simeq& \frac{1}{N}S_{\text{sing}}(i\pi-\theta)-\frac{1}{N^2}S_\text{sing}(\theta) \,. 
\eeqa
Note that the last term can be dropped inside the physical strip and its boundary \textit{except} at the upper boundary where both terms are small and of order $\cO(N^{-1})$. 
So, summarizing, we have 
\beqa
S_{\text{sing}}(\theta)&\simeq& \exp\Big( \frac{\theta}{i \pi} \log(N) \Big) \la{sum1}\\
S_{\text{sym}}(\theta)&\simeq&  \exp\Big(- \frac{\theta}{i \pi} \log(N)\Big) - \frac{1}{N^2} \exp\Big( \frac{\theta}{i \pi} \log(N) \Big) \la{sum2}\\ 
S_{\text{anti}}(\theta)&=& - S_{\text{sym}}(\theta) \la{sum3}
\eeqa
valid anywhere inside \textit{and} on the boundary of the physical strip. (But not elsewhere!)
At the upper boundary we get $S_{\text{sym}}(i\pi-\theta) \propto \frac{1}{N} \sin(\frac{\theta}{\pi} \log(N) )$ so we see that this solution exhibits an infinite amount of zeros (resonances) right at the upper boundary of the physical strip in the symmetric and anti-symmetric channels. We can again compare this expectation with the finite $N$ numerics to see that these zeros are there indeed, see also figure \ref{fig_maxSCA} below. These zeroes are a generic feature of the solutions in fact and arose already in another example, see figure \ref{fig_sheets1bsSING} above. The appearance of infinite resonances and periodicity along the real $\theta$ direction were encountered previously in other models like the ones studied in \cite{ZamZ4} and \cite{Mussardo}. In the context of the S-matrix bootstrap resonances were recently discussed in \cite{Doroud}.

Finally, note that while (\ref{sum1}-\ref{sum3}) are valid only in the physical strip, we can extend the solution to any value of $\theta$ since we can combine crossing symmetry with unitarity saturation which reads $S_\text{rep}(-\theta)=1/S_\text{rep}(\theta)$. 
For example, using unitarity we see that these infinitely many zeros at the upper boundary of the physical sheet become poles on the lower boundary of the so called mirror sheet $\text{Im}(\theta)\in[-\pi,0]$. Then we can use crossing to learn about the analytic structure of this solution in the strip above the physical strip where $\text{Im}(\theta)\in[0,2\pi]$ and then unitarity again etc. In this way, we can deduce the full analytic structure of this function to be as given as discussed in section~\ref{finitemaxSING} below. 
We see that at large $N$ the problem simplifies so much that we cannot only solve the maximization problem but also get key insights about the analytical structure of the various S-matrix elements. Indeed, assuming such structure persists at finite $N$ we will be able to guess the full analytic solution for this maximization problem below. 

\subsubsection{One bound state}
Next let us consider the problem of large $N$ theories with a single bound state whose coupling we maximize. 

First assume that the bound state is in the \textbf{singlet representation}. From the crossing relation (\ref{C1})
\beq
S_{\text{sing}}(i\pi-\theta)=
\frac{1}{N}S_{\text{sing}}(\theta)+\(\frac{1}{2}{\color{blue}-\frac{N}{2}}\)S_{\text{anti}}(\theta)+\({\color{blue}\frac{N}{2}}+\frac{1}{2}-\frac{1}{N}\)S_{\text{sym}}(\theta) \,, \la{C1Again}
\eeq 
we learn two things. The first one is that the t-channel pole residue is smaller than the s-channel pole by a factor of $1/N$ so that at large $N$ we can effectively drop it and keep a single pole at $\theta=i\lambda$ (the s-channel pole for a bound state of mass $m_\text{BS}=2 m \cos(\lambda/2)$). The second thing was discussed in the last section: to make the singlet component as large as possible it is optimal to anti-align the symmetric and anti-symmetric components; in that case the two terms in blue in \eqref{C1} add up so that the singlet component has maximal magnitude $N$ at the upper boundary. The punch-line is that we simply should multiply the solution (\ref{singletLargeN}) of the previous section by a function which introduces a single pole at $\theta=i\lambda$, is real for purely imaginary $\theta$ and is a phase in both boundaries of the physical strip. Such function is 
\beq
f_\lambda(\theta) \equiv 
\frac{1-e^{+i \lambda+\theta}}{e^{\theta}-e^{+i \lambda}} \,.
\eeq
We thus conclude that the optimal solution is given by
\beq
S_\text{sing}^\text{optimal} (\theta)=e^{\frac{\theta}{\pi i} \log(N)} \,f_\lambda(\theta) \la{1bsSINGlargeN}\,,
\eeq
%
%
%
and therefore
\beq
{\text{max}} \( g^2_\text{sing} \)=(2\sin\lambda) \times (2N^{\lambda/\pi}\,
\sin\lambda)\la{gsing_largeN}
\eeq
where the first factor is a simple Jacobian since we define the coupling as the residue in $s$. This indeed matches beautifully with the large $N$ extrapolation of our numerics as indicated in figure~\ref{fig_SCAlargeN}. We could also, as in the previous section, further determine the other channel S-matrices. The logic is exactly as before and leads to 
\beq
S_\text{sym}^\text{optimal}(\theta)=-S_\text{anti}^\text{optimal}(\theta)=\frac{1}{N}\[S^\text{optimal}_\text{sing}(i\pi-\theta)-\frac{1}{N}S^\text{optimal}_\text{sing}(\theta)\]\,. 
\eeq
At the boundary of the strip this solution exhibits again a rich pattern of resonances which one can indeed observe in the numerics at finite $N$. 

\begin{figure}
\centering
\includegraphics[scale=.91]{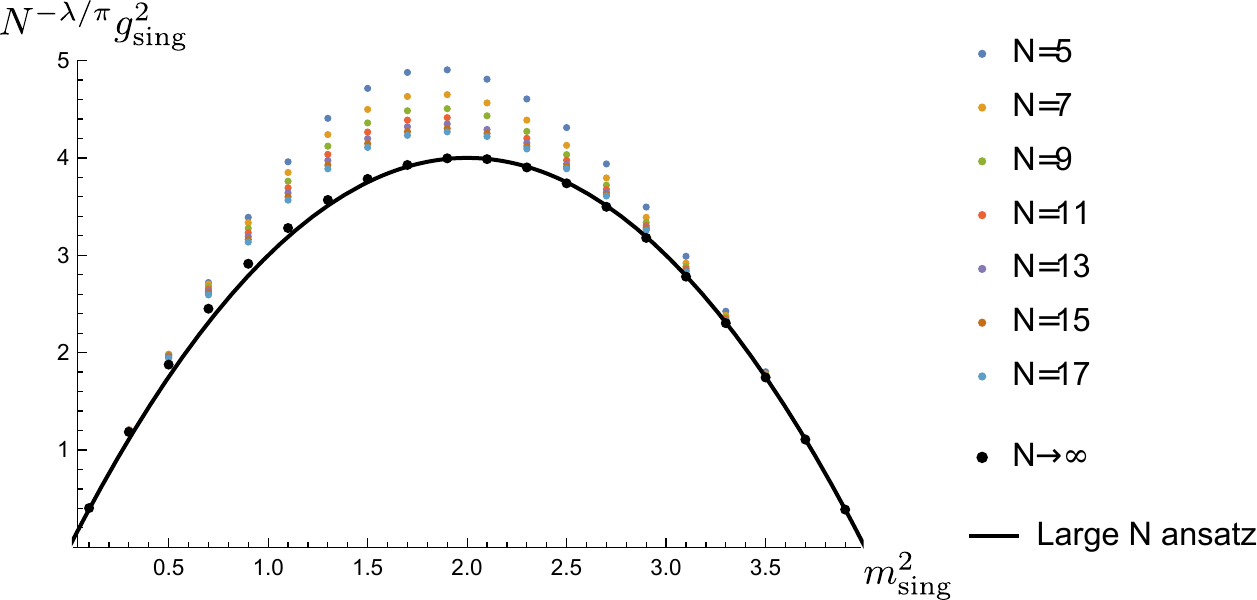}
\caption{Comparison between the numerics discussed in section~\ref{numNOint} and large $N$ analysis for the case with a single pole in the singlet representation. 
The maximum couplings are rescaled according to the power of $N$ in \eqref{gsing_largeN}. The finite $N$ results in figure~\ref{fig_onebs} (coloured points) are extrapolated to infinity (black points) and matched against the large $N$ ansatz (solid line).}
\la{fig_SCAlargeN}
\end{figure}

As a second example, consider a single pole in the \textbf{anti-symmetric representation} \textit{or} in the \textbf{symmetric representation}. From (\ref{C1Again}) we see that an $s$-channel pole in either of these representations induces a $t$-channel pole in the singlet component whose magnitude is~$N/2$ larger. In the right hand side of (\ref{C1Again}) the anti-symmetric and symmetric representations show up with opposite prefactors at large $N$ however the residue in the anti-symmetric case comes with a further minus sign -- recall (\ref{poles}). Therefore, we see that both problems lead to  the very same maximization problem as far as the singlet component is concerned: that of finding a $t$-channel pole in the singlet channel to be as \textit{large and negative} as possible. Following the same logic as in the previous case we thus conclude that the optimal solution for the singlet representation would read 
\beqa
S^\text{optimal}_\text{sing}(\theta)=- e^{\frac{\theta}{\pi i} \log(N)} \,f_{\pi-\lambda}(\theta)\,.
\eeqa
Computing the residue and multiplying by $-2/N$ and the usual Jacobian we then obtain the optimal large $N$ solution which as explained above coincides for these two problems: 
\beq
\left.\text{max} \(g^2_\text{anti}\)\right|_\text{single pole anti} =\left. \text{max} \(g^2_\text{sym}\)\right|_\text{single pole sym} =8 \sin^2(\lambda) N^{-\lambda/\pi} \la{gantisym_largeN}\,. 
\eeq
In figures \ref{fig_antisym_largeN}({\bf a}) and \ref{fig_antisym_largeN}({\bf b}) we see that the numerics for these two cases, although different at finite $N$, beautifully converge towards this universal prediction as we extrapolate them to infinite $N$. 

Note that in all three cases discussed in this section, even when we wanted to maximize couplings in other channels, the key was to study the singlet component at large $N$. This is in line with the intuition advocated at the beginning of this section. 

\begin{figure}
\centering
\includegraphics[width=\textwidth]{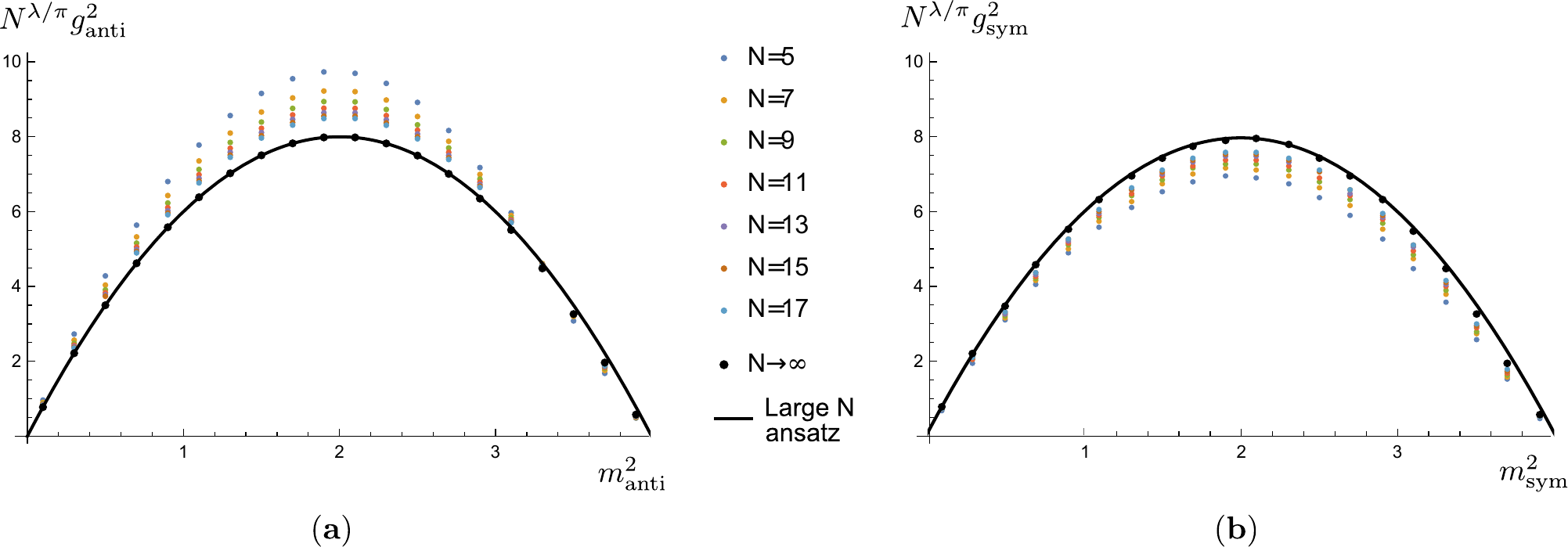}
\caption{Comparison between the finite $N$ numerics and the large $N$ analysis for ({\bf a}) a single bound state in the anti-symmetric representation and ({\bf b}) a single bound state in the symmetric representation.  As discussed in the main text, both cases lead to the same maximum coupling at large $N$ given in equation \eqref{gantisym_largeN} and represented here by the solid line.}
\la{fig_antisym_largeN}
\end{figure}

\subsubsection{Two bound states with same mass}
Let us consider a final large $N$ example where the theory has two bound states with equal mass, one in the {\bf singlet representation} and another one in the {\bf anti-symmetric representation} and we are maximizing the anti-symmetric coupling. This case is relevant since it contains the Gross-Neveu model for the particular value of the masses $m_\text{sing}=m_\text{anti}=2\cos(\pi/(N-2))$. As before, we want to find the large $N$ optimal solution for the singlet representation and use \eqref{C1Again} to read $g^2_\text{anti}$. This solution should be again a phase at the lower boundary of the physical strip and  of order $\cO(N)$ at the upper boundary. Since there is a bound state in the singlet representation, we should have an s- and a t-channel pole in that component. However, in contrast to \eqref{1bsSINGlargeN}, we cannot omit the t-channel pole because the anti-symmetric representation has also an s-channel pole, i.e. when evaluating~\eqref{C1Again} at $\theta=i\lambda$ we need to keep the first two terms on the right. We thus conclude that the optimal solution is given by\footnote{The product $f_\lambda(\theta) f_{\pi-\lambda}(\theta)$ is nothing but the usual CDD factor $f_\lambda(\theta) f_{\pi-\lambda}(\theta)=-\frac{\sinh\theta+i\sin\lambda}{\sinh\theta-i\sin\lambda}$. }
\beq
S^\text{optimal}_\text{sing}(\theta)=e^{\frac{\theta}{\pi i}\log(N)}\, \text{sign}(\lambda-\pi/2)  f_\lambda(\theta) f_{\pi-\lambda}(\theta)\la{singGNlargeN}\,.
\eeq
From \eqref{C1Again} we get the following large $N$ equation between the residues
\beq
\underset{\theta=i\lambda}{\text{Res}}\,S_\text{anti}(\theta)\approx\frac{2}{N}\[\underset{\theta=i\lambda}{\text{Res}}\,S_\text{sing}(\theta)+\frac{1}{N}\,\underset{\theta=i\pi-i\lambda}{\text{Res}}\,S_\text{sing}(\theta)\]\,,
\eeq
which after the replacement of the optimal solution \eqref{singGNlargeN} leads to the maximum coupling 
\beq
\max(g^2_\text{anti})=(2\sin\lambda)\times\(N^{-\lambda/\pi}-N^{-2+\lambda/\pi}\)4\lvert\tan\lambda\rvert\la{gGN_largeN}\,.
\eeq
The term $N^{-2+\lambda/\pi}$ in the previous equation is subleading for small $\lambda$ but becomes relevant as we reach $\lambda=\pi$ (e.g. as the mass of the bound states approaches zero). The divergence at $m^2_\text{BS}=2$ given by the screening of singlet and anti-symmetric poles is nicely reproduced by the factor of $\lvert\tan\lambda\rvert$ in \eqref{gGN_largeN}. In figure~\ref{fig_GNlargeN} we confirm that this result nicely matches with the large $N$ extrapolation of the numerics discussed in section~\ref{NumSec}. 

\begin{figure}
\centering
\includegraphics[scale=.9]{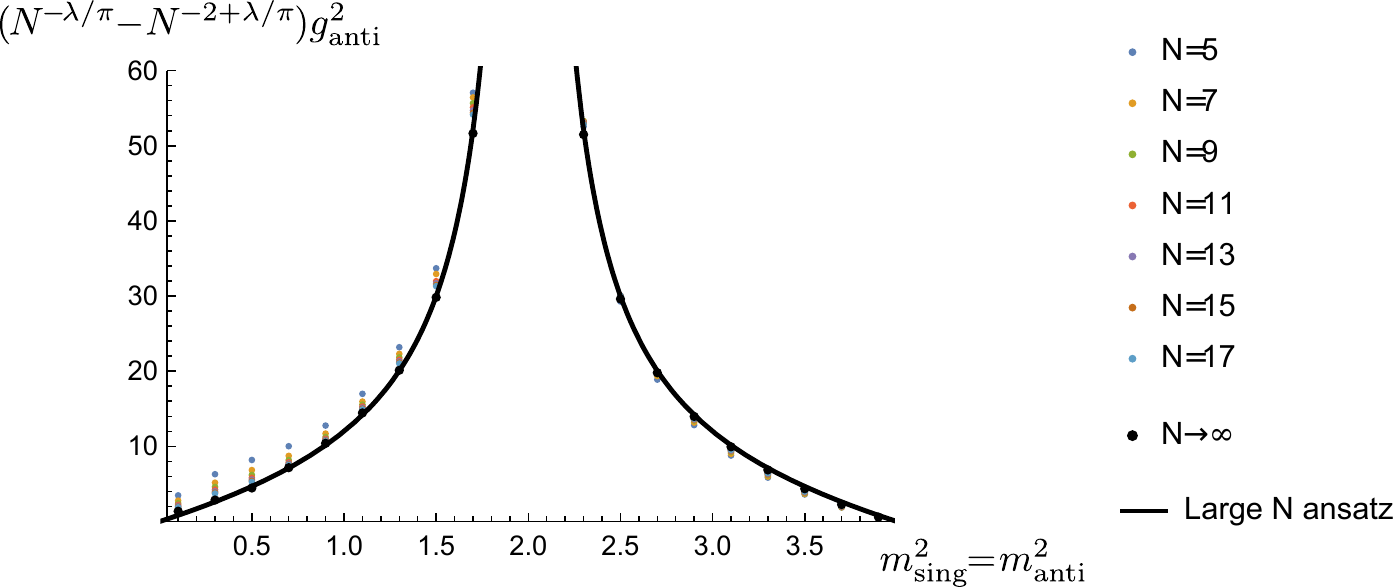}
\caption{Comparison between the finite $N$ numerics and the large $N$ analysis for two equal mass bound states in the singlet and anti-symmetric representations. The maximum couplings are rescaled by the power of $N$ in \eqref{gGN_largeN}. The finite $N$ results in figure~\ref{fig_GNline} (coloured points) are extrapolated to $N\rightarrow\infty$ (black dots) and compared to the large $N$ result  \eqref{gGN_largeN} (solid black line).}
\la{fig_GNlargeN}
\end{figure}

\subsection{Finite $N$} \label{finiteNsec}
In this section we discuss the analytic solutions at finite $N$ found in two cases: the maximization of $S_\text{sing}(i\pi/2)$ without bound states, and a single bound state in the singlet representation. As we will see, these solutions preserve some of the features found at large $N$ in the  previous section and nicely match the numerical results at finite $N$.
\subsubsection{Maximization/minimization singlet channel without bound states}\la{finitemaxSING}
Consider a theory without bound states where we maximize $S_\text{sing}(i\pi/2)$. As discussed in section~\ref{max_sing}, at large $N$ the symmetric and anti-symmetric representations antialign so that the singlet channel is as large as possible. Somewhat surprisingly, our numerics show that in the present case the equality $S_\text{anti}(\theta)=-S_\text{sym}(\theta)$  holds at finite $N$. We shall impose this condition in our derivation below. Another important feature observed at large $N$ is the presence of a sequence of zeros in $S_\text{sym}(\theta)$ at the upper boundary of the physical strip. For each of these zeros, unitarity+crossing implies there is an infinite sequence of zeros and poles in higher sheets. Since this is a general property of O($N$) S-matrices, let us carefully explain the logic leading to this structure.

Let us examine the consequences of a zero in the physical strip situated at some value $\theta^*$ for a given representation. From unitarity we see that there should be a pole in the $\theta\in[-i\pi,0]$ strip at $-\theta^*$ for the same representation.  Through the crossing equations, this pole implies the presence of more poles at  $i\pi+\theta^*$ for all representations in the  $\theta\in[i\pi,2i\pi]$ strip. We can use unitarity again to observe that all representations should have a zero at  $-i\pi-\theta^*$. Furthermore, crossing tells us that there are zeros as well at $2i\pi+\theta^*$.\footnote{It is easy to see in the crossing equations (\ref{C1}-\ref{C3}) that a single pole (zero) at $\theta^*$ in one of the representations does (not) imply extra poles (zeros) at $i\pi-\theta^*$. If we have more than one pole at $\theta^*$, the functions may conspire in such a way that their residues cancel so that there is no pole at $i\pi-\theta^*$ in a given representation. In contrast, zeros at $\theta^*$ in all representations does always imply zeros at $i\pi-\theta^*$ for all representations.} Continuing this logic we are naturally lead to the functions $F_a(\theta)$ defined in \eqref{SmatrixNLSM} and present in the integrable solutions of section~\ref{integrable}. These functions are explicitly unitary and encode some of the properties the S-matrices should satisfy from crossing.

Going back to the case at hand, we assume (based on numerical observations) that the zeros at $\theta\in i\pi+\mathbb R$ are periodically spaced by $\pi^2/\nu$, where $\nu$ is a parameter to determine. Following these zeros through crossing and unitarity as in the argument above we arrive at the following ansatz
\beq
{\bf S}(\theta)=
\begin{pmatrix}
\frac{\sinh\[\nu\(1-\frac{i\theta}{\pi}\)\]}{\sinh\[\nu\(1+\frac{i\theta}{\pi}\)\]}\\
-1 \\
1
\end{pmatrix}
\prod\limits_{n=-\infty}^\infty F_{\pi+\frac{in\pi^2}{\nu}}(-\theta)\la{maxSCA}\,.
\eeq
Replacing this ansatz into the crossing equations (\ref{C1}-\ref{C3}) fixes the remaining parameter $\nu$ to be $N=2\cosh(\nu)$. The analytic structure of this solution is described in figure~\ref{fig_maxSCA}.\footnote{For $N=2$ the solution \eqref{maxSCA} reduces to ${\bf S}(\theta)= \(\frac{\pi -i \theta }{\pi +i \theta },-1,1\)^\intercal F_{\pi}(-\theta)$.} 
\begin{figure}[h!]
\centering
\includegraphics[scale=1]{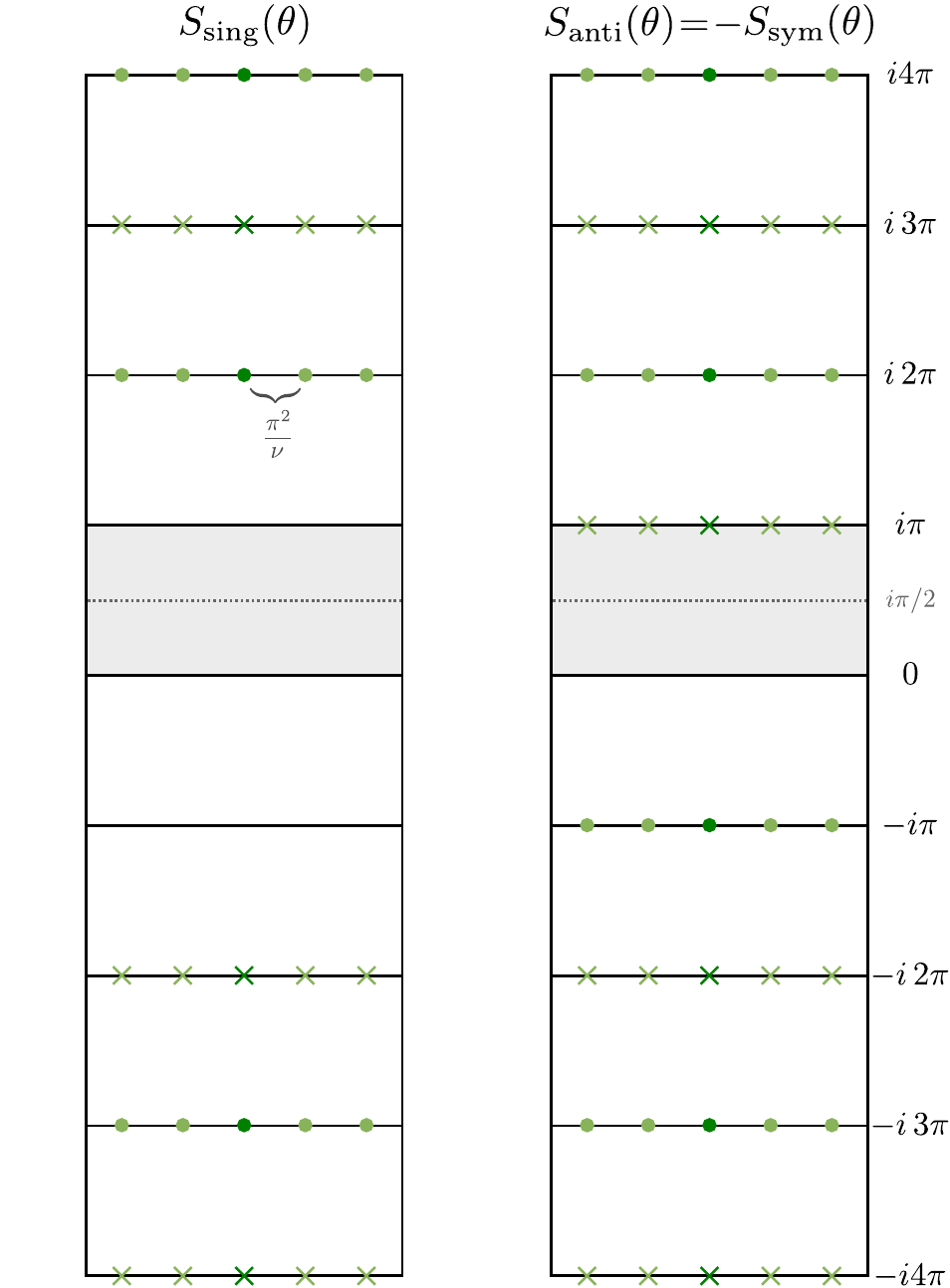}
\caption{Analytic structure of the solution to the maximization of the singlet representation $S_\text{sing}(i\pi/2)$ without bound states presented in \eqref{maxSCA}. The anti-symmetric and symmetric representations are the same up to a minus sign. The horizontal spacing between zeros and poles is $\pi^2/\nu$, where $\nu$ is given by $N=2\cosh(\nu)$. The central poles and zeros in dark green form the structure that remains in the $N=2$ limit. The minimization problem differs by an overall minus sign. }
\la{fig_maxSCA}
\end{figure}

Another way to write the S-matrices in \eqref{maxSCA} is
\beqa
S_\text{sing}(\theta)=e^{-i\theta\nu/\pi} \prod\limits_{l=1}^{\infty}	\frac{ \sinh\[\frac{\nu}{\pi}(i\theta-2l\pi)\]	\sinh\[\frac{\nu}{\pi}(i\theta+(2l+1)\pi)\] }{ \sinh\[\frac{\nu}{\pi}(i\theta+2l\pi)\]	\sinh\[\frac{\nu}{\pi}(i\theta-(2l+1)\pi)\]}\nn\\
S_\text{sym}(\theta)=-S_\text{anti}(\theta)=e^{i\theta\nu/\pi} \prod\limits_{l=1}^{\infty}\frac{\sinh\[\frac{\nu}{\pi}(i\theta+(2l-1)\pi)\]	\sinh\[\frac{\nu}{\pi}(i\theta-2l\pi)\]}{\sinh\[\frac{\nu}{\pi}(i\theta-(2l-1)\pi)\]	 \sinh\[\frac{\nu}{\pi}(i\theta+2l\pi)\]}\nn
\eeqa
which allows one to make direct contact with the large $N$ results of section~\ref{max_sing} rather trivially. Indeed, at large $N$ we have $\nu\simeq\log(N)$ so that all arguments in these expressions are very large. As such, inside each $\theta$ strip we can simplify all $\sinh$'s to one of its exponentials. In this way we see that inside any $\theta$ strip any component is given by a simple power of $N$ times $e^{\mp i\theta\nu/\pi}$. For instance, in the physical sheet we recover (\ref{sum1}) and (\ref{sum2}) (recall that inside the strip we can drop the last term in (\ref{sum2})). 
The above representation is also more suitable for numerical evaluation, since when truncating the product to some $l_\text{max}$ we keep the poles and zeros closer to the physical strip. In fact, it was using this representation with $l_\text{max}=50$ that we generated the yellow curve in figure~\ref{figSsingletMiddleStrip}.

If instead we want to minimize $S_\text{sing}(i\pi/2)$ we would be led to the same solution with an overall minus sign. 

Given that the solution~\eqref{maxSCA} saturates unitarity, it is natural to ask if it also obeys Yang-Baxter equations. As one can check it does satisfy factorization and in fact was written before in the appendix of \cite{newYB}!\footnote{There, equation~A.5 should have an infinite product instead of a sum.} To our knowledge, a physical model corresponding to this integrable S-matrix is yet to be identified.\footnote{This integrable S-matrix bears strong resemblance with the one in \cite{Zamolodchikov:1990dg} where also $S_\text{anti}(\theta)=-S_\text{sym}(\theta)$ (or equivalently $\sigma_2(\theta)=0$ in the decomposition \eqref{SmatrixON}). In that case there is also a periodicity related to $N$, which takes values $\lvert N\rvert<2$ and for $N=0$ this solutions beautifully relates to self-avoiding polymers, see also \cite{ Smirnov:1991ew,Fendley:2001ez}. The relative coefficients of the various S-matrix channels in \cite{newYB} and \cite{Zamolodchikov:1990dg} differ by a simple analytic continuation $\nu\to i \mu$ while the overall factor is related in a more involved way. It would be interesting to explore further the connections between these two solutions.}

\subsubsection{One bound state in singlet representation}
Let us now derive the finite $N$ solution for the case of a theory with a single bound state in the {\bf singlet representation}. 

The first thing we notice from the numerics is that, as in the previous example, the large $N$  equality $S_\text{anti}(\theta)=-S_\text{sym}(\theta)$ holds at finite $N$. Secondly, there are also zeros at the upper boundary of the physical strip. This time however there is no zero at $\theta=i\pi$ and what was a single sequence of zeros in the previous example splits now into two strings starting at $\theta=i\pi\pm\zeta$, where $\zeta$ is a parameter to determine. In the following we will assume that $S_\text{anti}(\theta)=-S_\text{sym}(\theta)$ and that the spacing between zeros in each string is again given by $\pi^2/\nu$ which is consistent with numerical observations (see green structure in figure~\ref{fig_1bsSCA}).\footnote{The fact that the there are two copies of strings of zeros at $\theta=i\pi\pm\zeta\pm n\pi^2/\nu$ follows from real analyticity $S^*(s)=S(s^*)$ or $S^*(\theta)=S(-\theta^*)$.}

Let us now consider the implications of a single pole at $\theta=i\lambda$ in the singlet channel corresponding to the bound state. For this we use the crossing relations simplified by the condition $S_\text{anti}(\theta)=-S_\text{sym}(\theta)$. One of the equations reads
\beq
S_\text{sym}(\theta)=\frac{1}{N}\[S_\text{sing}(i\pi-\theta)-S_\text{sym}(i\pi-\theta)\]\,.
\eeq
If we evaluate this equation at $\theta=i\lambda$, we conclude that the t-channel poles should cancel each other so that $S_\text{sym}(i\lambda)$ remains finite (recall that there is only one s-channel pole for the singlet representation). In other words, their residues must be the same. One can check using unitarity that this implies the existence of a double zero (pole) for $S_\text{sym}(2i\pi-i\lambda)$ ($S_\text{sym}(-2i\pi+i\lambda)$).\footnote{We have that $S_\text{sing,sym}(\theta)\sim\frac{r}{\theta-(i\pi-i\lambda)}$ and $S_\text{sym}(i\pi+\theta)=\frac{1}{N}\[\frac{1}{S_\text{sing}(-\theta)]}-\frac{1}{S_\text{sym}(-\theta)}\]$ so that $S_\text{sym}(2i\pi-i\lambda)\approx\frac{1}{N}\[\frac{1}{r}(\theta-i\pi+i\lambda)-\frac{1}{r}(\theta-i\pi+i\lambda)\]$ which indeed has a double zero.} 
By the crossing+unitarity logic described before, the double pole will propagate to higher sheets creating the pattern depicted in blue in figure~\ref{fig_1bsSCA}. Let us stress the general property of the O($N$) S-matrices we just derived: whenever a pole appears only in the singlet representation there will be a tower of double poles and zeros in higher sheets for all representations.

\begin{figure}[t!]
\centering
\includegraphics[scale=1]{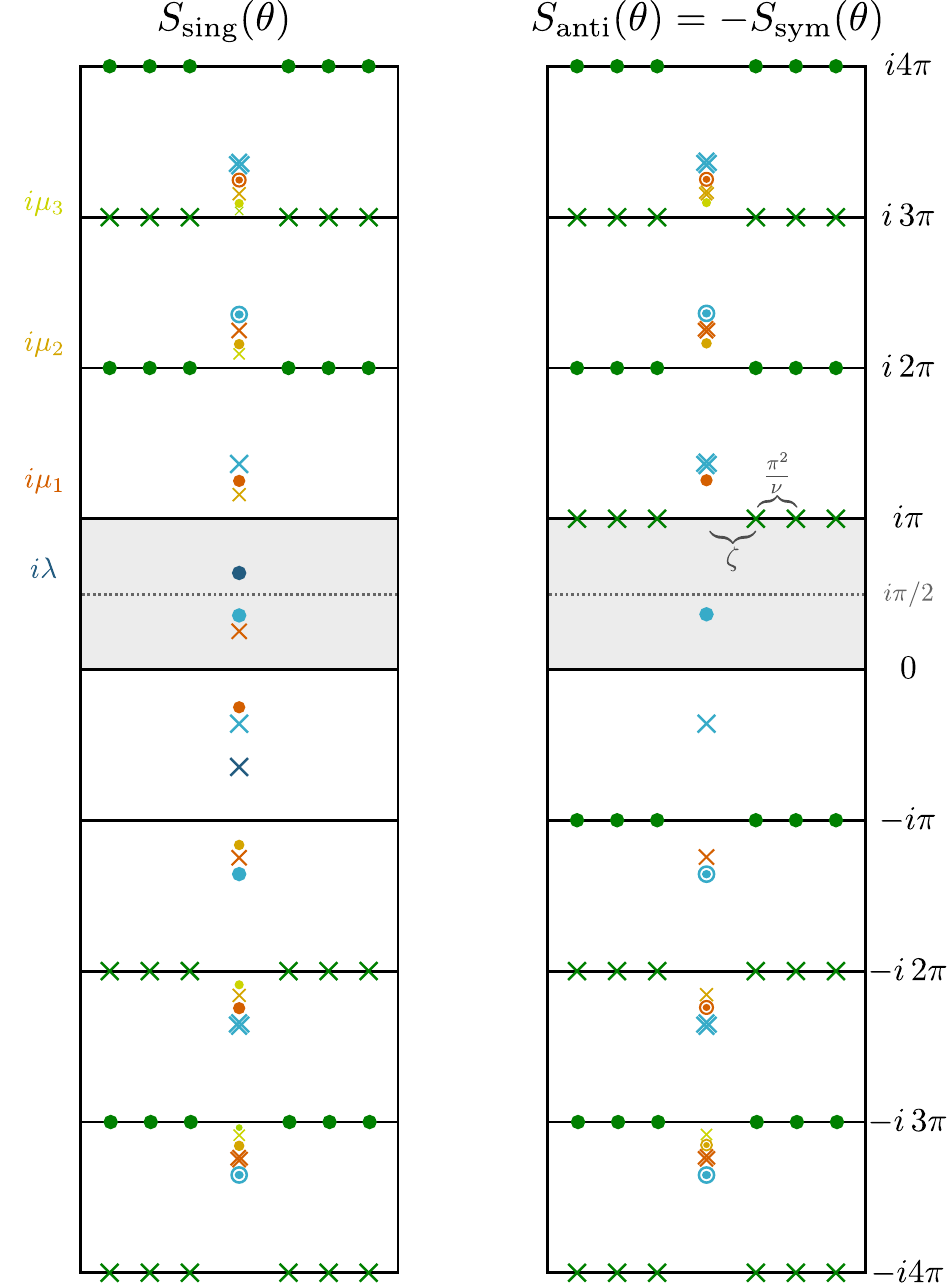}
\caption{Analytic structure of the solution to the maximization of $g_\text{sing}$ with no other bound states given in \eqref{S1bsSCA}. The anti-symmetric and symmetric representations are the same up to a minus sign. The double crosses $\TextVCenter{\protect\dzero}$ and circles $\TextVCenter{\protect\dpole}$ represent double zeros and poles, respectively. The parameters $\mu_i$ specify the position of the new zeros and poles appearing in the singlet representation. The first zeroes at the upper boundary of the physical strip depicted in green are located at $\theta=i\pi\pm\zeta$ and the rest are spaced by multiples of $\pi^2/\nu$. }
\la{fig_1bsSCA}
\end{figure}

The next thing to note is that the value ${\bf S}(0)$ fixes ${\bf S}(i\pi n)$ for any $n\in\mathbb Z$ through crossing. In particular, if we read from the numerics the value ${\bf S}(0)$ we can compute the sign of the S-matrices at ${\bf S}(i\pi n)$. In the present case we have ${\bf S}(0)=(1,1,-1)^\intercal$ which leads to alternating signs every $i\pi$ for all representations. For $S_\text{sing}(\theta)$, this means that as we move in the imaginary line segment $\theta\in[0,i\pi]$, we should start with a positive function which then has the s- and t-channel poles and end with a negative value. Since each pole changes the sign of the function, we conclude that there should also be a zero in this line (another pole would imply there are other bound states) at some position $\theta=i\mu_1-i\pi$. For  $S_\text{anti}(\theta)=-S_\text{sym}(\theta)$ the single t-channel pole is consistent with the change of sign. Following the new zero through crossing and unitarity --keeping in mind the double poles and zeros produced in higher sheets-- we arrive at the orange structure in figure~\ref{fig_1bsSCA}. 

If we repeat the exercise in the second strip $\theta\in[i\pi,2i\pi]$ we easily see the need for a new zero in $S_\text{sing}(\theta)$ (yellow cross in figure~\ref{fig_1bsSCA}). This is actually the case in every strip, so that we end up with infinite parameters $\mu_n$ which label the position of the first new pole in the $n-$th strip. 

\begin{table}[t]
\centering
\begin{tabular}{l l}
\hline
$\zeta$ &1.804\\
$\mu_1$ & 5.401\\
$\mu_2$ & 8.548\\
$\mu_3$ & 11.68\\
$\mu_4$ & 14.86\\
\hline
\end{tabular}
\caption{Values for the parameters $\zeta$ and $\mu_{i\leq4}$ with $N=5$ and $m_\text{sing}^2=3.1$ ($\lambda=0.98$), obtained by the method explained in appendix~\ref{appmus}.
}
\label{tab:mus}
\end{table}

Collecting the analytic structure discussed above and depicted in figure~\ref{fig_1bsSCA}, we write the ansatz for the S-matrix
\beqa
{\bf{S}}(\theta)&=&
\begin{pmatrix}
G(\theta)\\
1 \\
-1
\end{pmatrix}
\;\frac{\pi-\lambda+i \theta }{\pi-\lambda-i \theta }\,F^2_{\pi-\lambda}(\theta)
\( \prod\limits_{i=1}^{\infty}\frac{\mu_i +i \theta}{\mu_i  -i \theta}\,F^2_{\mu_i }(\theta) \) 
\( \prod\limits_{n=0}^{\infty}F_{-i\zeta-\frac{in\pi^2}{\nu}}(\theta) F_{i\zeta+\frac{in\pi^2}{\nu}}(\theta) \)\,,\nn\\ 
\la{S1bsSCA}
\eeqa
where $\mu_i$ and $\zeta$ are parameters to determine in function of $\lambda$ and $N$ and we have defined
\small
\beq
G(\theta)=\frac{i\theta-\lambda}{i\theta+\lambda}\,\frac{i\theta+\lambda-2\pi}{i\theta-\lambda+2\pi}\(\prod\limits_{i=1}^\infty\frac{i\theta+\mu_i-\pi}{i\theta-\mu_i+\pi}\,\frac{i\theta-\mu_i-\pi}{i\theta+\mu_i+\pi}\)\frac{\Gamma\[\tfrac{\nu}{\pi^2}(\theta-i\pi+\zeta)\]\Gamma\[\tfrac{\nu}{\pi^2}(-\theta+i\pi+\zeta)\]}{\Gamma\[\tfrac{\nu}{\pi^2}(-\theta-i\pi+\zeta)\]\Gamma\[\tfrac{\nu}{\pi^2}(\theta+i\pi+\zeta)\]}\,.
\eeq
\normalsize
The remaining parameters $\mu_i$ and $\zeta$ can be fixed by using crossing symmetry. As explained in details in appendix~\ref{appmus}, the system of equations we need to solve has the nice form
\beq
\[\prod\limits_{i=1}^{\infty}\frac{\mu _i^2-\pi ^2 n^2}{\mu _i^2-\pi ^2 (n+1)^2}\] h(n,\zeta)=1\,,
\la{eqmus0}
\eeq
which can be solved numerically very efficiently and to arbitrary accuracy. An example of the values obtained by solving the above equations is given in table~\ref{tab:mus}. We suspect \eqref{eqmus0} should have a nice and simple physical interpretation; it would be very interesting to find it.

The cases discussed above satisfied $S_\text{anti}(\theta)=-S_\text{sym}(\theta)$ which greatly simplified our task of finding a solution to unitarity and crossing equations. For all other cases we studied this equality no longer holds, so that finding the full analytic solution is more complicated. Although we do not have the analogue of \eqref{eqmus0} for other cases, we were able to obtain (at least some of) the analytic structure of other solutions. For example, considering bound states in the {\bf singlet and anti-symmetric representations} with the same mass, one finds --for the mass range $m_\text{sing}=m_\text{anti}<m_\text{GN}$-- the poles and zeros depicted in figure~\ref{fig_exotic}. This case exhibits an ever richer structure with pairs of zeros inside the physical strip
and patterns that repeat each other in a fractal way as we move outside the physical strip. Other maximization problems lead to S-matrices of roughly the same complexity. It would be nice to obtain the full analytic solution in these cases.

\begin{figure}[t!]
\centering
\includegraphics[scale=1]{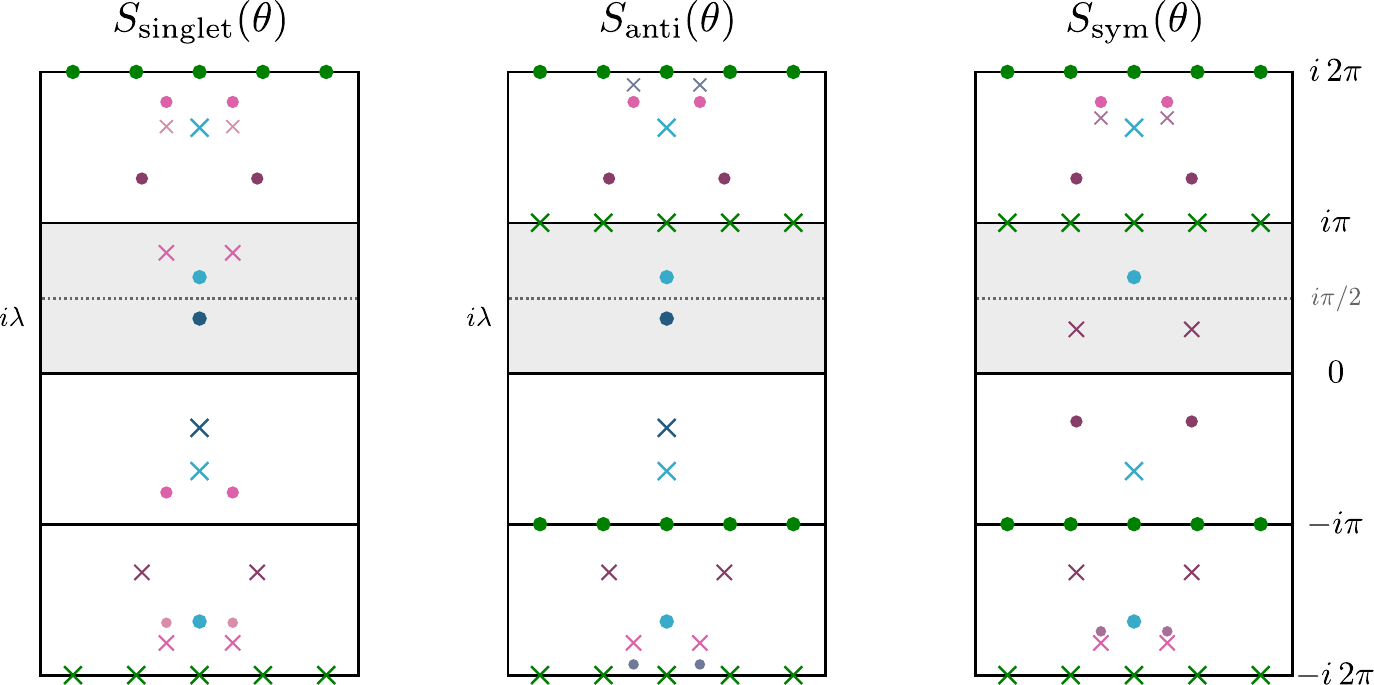}
\caption{Sketch of the analytic structure of more exotic S-matrices found by considering bound states in the singlet and anti-symmetric representations with $m_\text{sing}=m_\text{anti}<m_\text{GN}$ and maximizing the anti-symmetric coupling $g_\text{anti}$.}
\la{fig_exotic}
\end{figure}

\section{Discussion}\la{discussion}

We analysed the space of O($N$) symmetric two-dimensional relativistic quantum field theories whose lightest particles transform in the fundamental representation. We carved out part of this space by looking for $2\to 2$ candidate S-matrices which maximize various couplings (be them cubic couplings given by the physical residues of the S-matrix or effective four point couplings given by the S-matrix amplitudes evaluated in some symmetric point in the physical sheet).

In the boundary of this theory space we identified known integrable S-matrices describing the excitation scattering in the O($N$) non-linear sigma model (NLSM), the O($N$) Gross-Neveu model, the kinks of the Sine-Gordon theory or the yet unknown model for solution~\eqref{maxSCA} first discovered by Hortacsu, Schroer and Thun in \cite{newYB}. The NLSM S-matrix was beautifully identified to lie at a cusp of the space of unitarity and crossing invariant S-matrices in \cite{Martin}. It would be very interesting to connect our various bounds to their proposed picture for the S-matrix space and its cusps. We studied, for instance, the other cusp found in that work and found that its analytic structure is reminiscent of the general pattern found here, with infinitely many resonances at the boundary and inside the physical sheet. 

We can already anticipate that the less known and quite exotic integrable solution~\eqref{maxSCA} is also a cusp. This is yet another reason to look for the corresponding physical theory. We expect it should describe some planar processes as the self-avoiding polymers of \cite{Zamolodchikov:1990dg} since the only allowed color structures are
\beq
\mathbb{S}_{ij}^{kl} (s)\;=\; \sigma_1(s)\;\vcenter{\hbox{\includegraphics{K.pdf}}}\quad+\quad\sigma_3(s)\;\vcenter{\hbox{\includegraphics{P.pdf}}} \,.\nn
\eeq
It would be very interesting to explore the possible connections with the models described in \cite{  Smirnov:1991ew} and \cite{Fendley:2001ez}. 

So far, we encountered already at least three cusps  in the space of theories where integrability holds: The NLSM found in \cite{Martin}, the free theory of section~\ref{sec3int} and the more exotic solution of \cite{newYB} in section~\ref{finitemaxSING}. It would be fascinating to understand mathematically the emergence of Yang-Baxter equations at the cusps of this space. While natural from a physical point of view, the emergence of these cubic equations remains a completely mathematical mystery in this context. Can we find other solutions to YB (perhaps some never discovered before) for other symmetry groups?

Also at the boundary of this physical space we identified even more exotic S-matrices. These saturate unitarity so that \beq
\mathbb{S}_{2\to 2}^\text{exotic}(s)=\sum_{\text{rep}} S_\text{rep}(s) \mathbb{P}_\text{rep}  \qquad \text{with} \qquad |S_\text{rep}(s)|^2=1 \qquad \text{for}\qquad s>4m^2 \,.
\eeq
where the sum is over the three possible representations $\text{rep}=\(\text{singlet},\text{anti},\text{sym}\)^\intercal$. These theories have an extremely rich pattern of resonances with infinitely many of them in the various sheets of the Mandelstam plane as explored in the main text (see for example figures~\ref{fig_1bsSCA} and \ref{fig_exotic}).\footnote{The resonances come about from an emergent periodicity in hyperbolic rapidity reminiscent of \cite{Zamolodchikov:1979ba}. As pointed out in this paper, such periodicity in $\theta \sim \log(s)$ seems to point towards some sort of limit cycle behavior in the UV. The later are ruled out in local two dimensional unitary field theories \cite{cTheorem}. It would be very interesting to find an S-matrix theoretical version of the c-theorem which could rule them out more generally. Perhaps such argument could shed light on the physics of the S-matrices encountered here. We thank Sasha Zamolodchikov for interesting discussions on this point and for bringing \cite{Zamolodchikov:1979ba} to our attention.} Because they saturate unitarity, these optimal solutions admit zero particle production so that $\mathbb{S}_{2\to 3}^\text{exotic}(s)=\mathbb{S}_{2\to 4}^\text{exotic}(s) = \dots = 0$ just like the integrable models. However, contrary to those these exotic S-matrices do \textit{not} obey the Yang-Baxter relations. 

Can we have proper physical theories with no particle production and no Yang-Baxter? Probably not. Since Yang-Baxter is not satisfied,  the $3\to 3$ S-matrix cannot consistently factorize into a product of two-to-two S-matrices (like for an integrable theory). If this S-matrix is not factorized (and hence supported on an additional conservation $\delta$ function consistent with this factorization), it should have a regular part. This regular part can be analytically continued by crossing into $S_{2\to 4}$ which thus should not be strictly zero \cite{Iagolnitzer,Iagolnitzerbook}. These exotic S-matrices, with strictly zero particle production are thus likely unphysical. 

Having said that, perhaps we need not completely discard these S-matrices altogether. After all, while we can probably not accept a strictly zero particle production in a theory whose two body S-matrix does not obey the factorization conditions, can this particle production be very small? In other words, could these S-matrices, with their very rich analytic properties and their infinitely many resonances, be close enough to real physical S-matrices which simply have little particle production? 

To properly analyze this question we could try to bootstrap not only the two body S-matrix but also multi-particle components, such as $S_{2\to 4}$ and $S_{3 \to 3}$, altogether and figure out the fate of these S-matrices in this more complete setup. This is a fascinating but very involved problem which we are currently attacking together with Alexandre Holmrich. Further intuition about the exotic S-matrices might be found in the study of the "special form factors" introduced in \cite{Smirnov} properly generalized to a setup with O($N$) symmetry. What will these tell us about deformations around the NLSM or the Gross-Neveau model for instance? Can this shed light over the exotic S-matrices we found for masses near those of the Gross-Neveu model?

In the meantime, we can preform some phenomenological games and add some particle production by hand. More precisely, we can redo the numerics modifying the unitarity constraints to read $ |S_\text{rep}(s)|^2 \le f_\text{rep}(s)$ where $f_\text{rep}(s)=1-\sigma_\text{rep}^{(\text{abs})}$ is equal to $1$ until we reach inelastic thresholds and decays to zero for larger values of $s$. Of course we do not know what these functions are so the best we can do  we can do in this simple exercise is to set them to some arbitrary simple functions with these properties and see what one gets. If all functions are taken to be the same (for example, we could set $f_\text{rep}(s)=\exp(-\alpha (s-16m^2)^2 \theta(s-16m^2))$ starting at the 4 particle threshold) then the new optimization problems are actually trivially related to the ones in the main text as $S_\text{with particle production}(\theta)=F(\theta) S_\text{main text}(\theta)$ where $F(\theta)$ is given by formula (11) in \cite{Paper2}. In particular, because the solutions are simply dressed by this overall function, all the rich pattern of zeros and poles which we found are still present with particle production.  More realistically, different channels will produce particles at different rates so things will be less trivial. At large $N$ we can repeat most of the analysis in section~\ref{largeNsec} for fixed imposed particle production. We can also easily run numerics with modified unitarity. The conclusion of this simple analysis -- using some randomly chosen particle production functions -- is that the various resonances are  still there but their positions move around slightly; some acquire some imaginary part, some stay on the $t$-channel cut. For low energy the S-matrices are comparable to the unphysical ones we got without any particle production provided particle production is not large or changes too abruptly. 
This is perhaps not so surprising since particle production is a high energy effect of sorts while our maximization probes mostly the low energy region of the S-matrices. 

\textit{If} the infinitely many resonances present in the integrable solution \eqref{maxSCA} and more exotic S-matrices are indeed there for some physical theory (albeit in some slightly shifted positions in the second case), it would be fascinating to unveil their physical origin.

\section*{Acknowledgements}
We thank Benjamin Basso, Freddy Cachazo, Frank Coronado, Patrick Dorey, Yifei He, Alexandre Homrich, Martin Kruczenski, Miguel Paulos, Joao Penedones, Jon Toledo, Balt van Rees, and specially Alexander Zamolodchikov for numerous enlightening discussions and suggestions. Research at the Perimeter Institute is supported in part by the Government of Canada through NSERC and by the Province of Ontario through MRI. 
This research received funding from the grant CERN/FIS-NUC/0045/2015. This work was additionally supported by a grant from the Simons Foundation (PV: \#488661). PV thanks FAPESP grant 2016/01343-7 for partial financial support. LC would like to thank CONACyT and ERC Starting Grant 679278 Emergent-BH for partial financial support.

\appendix
\section{Fixing parameters in analytic solution}\la{appmus}

In this appendix we explain how to compute the parameters $\mu_i$ and $\zeta$ in the analytic solution \eqref{S1bsSCA}. For this, it is convenient to look at the product $S_\text{anti}(\theta)S_\text{anti}(\theta+i\pi)$ which has a much simpler analytic structure. In particular one can compute explicitly the second infinite product in \eqref{S1bsSCA} and get 

\beqa
S_\text{anti}(\theta)S_\text{anti}(\theta+i\pi)&=&
\[\prod\limits_{i=1}^{\infty}\frac{\mu _i^2+\theta^2}{\mu _i^2+(\theta+i\pi)^2}\] 
\frac{\theta ^2+(\pi -\lambda )^2}{(\theta -i \lambda +2 i \pi ) (\theta +i \lambda )}\times\nn\\
&&\times\,
\frac{\Gamma \left[\frac{\nu }{\pi ^2}(-i \pi +\zeta -\theta ) \right] \Gamma \left[\frac{\nu }{\pi ^2}(i \pi +\zeta +\theta ) \right]}{\Gamma \left[\frac{\nu }{\pi ^2}(\zeta -\theta ) \right] \Gamma \left[\frac{\nu }{\pi ^2}(\zeta +\theta ) \right]}\,.\la{prodtheta}
\eeqa

Using ${\bf S}(0)=\(1,1,-1\)^\intercal$, one can then compute the values of ${\bf S}(i\pi n)$ ($n\in\mathbb Z$) using crossing and unitarity to derive
\beqa
S_\text{anti}(i\pi n)S_\text{anti}(i\pi(n+1))=\frac{\cosh\[n\,\nu\]}{\cosh\[(n+1)\nu\]}\la{effmus}\,.\la{effmus}
\eeqa

Evaluating the product \eqref{prodtheta} at $\theta=i\pi n$, one can rearrange terms in \eqref{effmus} and arrive at a more suggestive form, reminiscent of Bethe equations for $\mu_i$:
\beq
\[\prod\limits_{i=1}^{\infty}\frac{\mu _i^2-\pi ^2 n^2}{\mu _i^2-\pi ^2 (n+1)^2}\] h(n,\zeta)=1\,,\la{eqmus}
\eeq
where we have defined $h(n,\zeta)$ by
\beq
 h(n,\zeta)=\frac{(\lambda -(n+1)\pi ) (\lambda +(n-1)\pi  )}{(\lambda  -(n+2) \pi ) (\lambda +\pi  n)}
\frac{\Gamma \left[\frac{(\zeta -i (n+1) \pi ) \nu }{\pi ^2}\right] \Gamma \left[\frac{(i \pi  (n+1)+\zeta ) \nu }{\pi ^2}\right]}
{\Gamma \left[\frac{(\zeta -i n \pi ) \nu }{\pi ^2}\right] \Gamma \left[\frac{(i \pi  n+\zeta ) \nu }{\pi ^2}\right]}
\frac{\cosh[\nu  (n+1)]}{\cosh(\nu\,  n)}\,.
\eeq
Given $N=2\cosh\nu$ and a bound state mass $m_\text{sing}$ (or equivalently $\lambda$), there is an infinite number ($n\in\mathbb Z$) of coupled equations \eqref{eqmus} for infinite variables $\mu_i$ and $\zeta$. In practice, we can truncate the number of parameters and equations, assuming that the remaining $\mu$'s are spaced by $\pi$. That is, we truncate the infinite product in \eqref{eqmus} to some $i_\text{max}$ and assume $\mu_{i+1}-\mu_i=\pi$ for $i\geq i_\text{max}$. 
This approximate spacing is observed numerically and can be understood from the fact that as we go further away from the physical strip we have more zeros and poles corresponding to previous $\mu$'s and less space between these and the boundary of the of the corresponding strip (see for example the $\theta\in i[3\pi,4\pi]$ strip in figure~\ref{fig_1bsSCA}). The truncation leads to the simple modification
\beq
\[\prod\limits_{i=1}^{i_\text{max}}\frac{\mu _i^2-\pi ^2 n^2}{\mu _i^2-\pi ^2 (n+1)^2}\] \frac{\mu _{i_{\max }}+\pi  (n+1)}{\mu _{i_{\max }}-\pi  n}h(n,\zeta)=1\,.\la{effeqmus}
\eeq

Now we only need to consider $i_\text{max}+1$ different values of $n$ to solve the system of equations. The values obtained in table~\ref{tab:mus} were found by evaluating the effective equations \eqref{effeqmus} for $n=1,\ldots,5$ and $i_\text{max}=4$. We performed this exercise for various $m_\text{sing}$ and $N$ and found always perfect agreement with the position of zeros and poles in our numerics.

\end{document}